\pdfoutput=1

\newcommand{\degr}{$^\circ$}
\newcommand{\caproman}[1]{\uppercase\expandafter{\romannumeral#1}}

\newcommand{\BGsite}{http://users.physics.harvard.edu/\raisebox{-4pt}{$\tilde{\;\;}$}gottschalk}
\newcommand{\PTCOGsite}{http://ptcog.web.psi.ch/}
\newcommand{\BGmail}{bgottsch\,@\,fas.harvard.edu}

\documentclass{article}

\title{\bf Comprehensive proton dose algorithm\\using pencil beam redefinition\\and recursive dynamic splitting}
\author{B. Gottschalk\thanks{Harvard University Laboratory for Particle Physics and Cosmology, 18 Hammond St., Cambridge, MA 02138, USA, \BGmail}}
\date{\today}

\usepackage{amsmath}
\usepackage{latexsym}
\usepackage{graphicx}

\begin{document}

\maketitle
\vspace{.5in}
\begin{center}{\large\bf For Andy Koehler (1930--2015), friend and mentor.}\end{center}

\clearpage
\begin{abstract}
\noindent
We describe a new pencil beam (PB) algorithm that computes, from first principles, absolute fluence or dose per incident proton charge in a known heterogeneous terrain irradiated by known proton beams. PB transport is by standard Fermi-Eyges theory. Heterogeneites are handled by breaking up PBs using redefinition and recursive dynamic splitting.

The terrain is discretized into uniform or heterogeneous slabs by $z$-planes perpendicular to the nominal beam direction. In each slab, heterogeneity boundaries (if any) are either simple geometric shapes or, more generally, polygons. All objects in the terrain, including collimators, are represented in that fashion. There are no special `beam limiting devices'.

The charge carried by a PB is represented by an integer $i_g$ (how many generations the PB is removed from the parent) and a continuous variable $q$, such that total charge $=q\times0.5^{i_g}$\,nC. The longitudinal variable, rather than kinetic energy or residual range in water, is $pv$ (proton momentum times speed). Other PB parameters are conventional: central axis position and direction and three Fermi-Eyges moments representing PB size, divergence and emittance. (Non cylindrical incident beams would require three more.) 

PB parameters are defined only on $z$-planes. The actual `ur-beams' irradiating the terrain, and the virtual PBs generated during execution, are parameterized and processed identically. 

Each material in the terrain is treated on its own merits. There is no reference material (e.g. water) and concepts such as `radiological path length' are not used. The {\O}ver{\aa}s approximation $R=a(pv)^b$, fitted to standard range-energy tables, is used to compute the diminution of $pv$, and an accurate scattering power $T_\mathrm{dM}(pv)$ represents multiple Coulomb scattering. Thus, non water-like materials (e.g. beam line components, surgical implants) are handled accurately.

The basic program, a PB stack processor, nested in a loop over ur-beams, nested in a loop over states of the terrain, is equally suited to scattered beams or pencil beam scanning. In a scattered beam, the terrain is more complicated and has more states (e.g. modulator steps) and there are far fewer ur-beams (typically one per state). In pencil beam scanning, the reverse is true.

The PB stack processor examines a PB at a $z$-plane. Depending on various factors (e.g. proximity to a heterogeneity) the PB may be redefined (replaced on the stack by a horde of new, smaller PBs) or split (replaced by seven new PBs) or transported to the next $z$-plane (PB parameters replaced by new values). After transport, if the new $z$-plane is a designated measuring plane (MP), the dose from the current PB at each point of interest is added in, and the loop repeats. When a PB ranges out or reaches the last $z$-plane in the problem, it is deleted and the next PB is examined at the first $z$-plane. When the last PB is deleted, control passes out and the next ur-beam is placed on the stack. After the last ur-beam is processed, control passes to the next state of the terrain.

PB redefinition is more or less conventional: a PB is replaced by many smaller PBs arranged in a hexagonal array in such a way that the phase space parameters and charge of the mother PB are conserved. PB splitting, by contrast, replaces the mother by a central daughter carrying one quarter of the charge and traveling in the same direction, and six daughters arranged in hexagonal array, radiating from the virtual point source of the mother, each with one eighth of the charge, the whole conserving emittance.

In common with other PB algorithms we ignore secondaries from hard scatters, covering only those protons (typically 80\%) that stop by multiple EM interactions with atomic electrons. Two extremes are covered, however, in choosing the effective mass stopping power used in converting fluence to dose. If the contribution of secondaries in a particular problem is thought to be negligible, an electromagnetic mass stopping power $S_\mathrm{em}/\rho$ in the `dose-to' material may be computed from first principles. If, by contrast, the region of interest is within a broad beam where transverse equilibrium (including secondaries) prevails, a mixed mass stopping power $S_\mathrm{mixed}/\rho$ may be derived from an experimental broad-beam Bragg peak or an integral depth-dose (IDD) measurement. In that case the `dose-to' material will probably be water.

After describing the algorithm, with intermediate tests, in great detail, we present three examples. The first two involve collimator scatter, solved for the first time (excepting Monte Carlo) by this work, and a stringent test of the algorithm. The third compares this work with the widely used Hong algorithm. We conclude by discussing how the present work might be put into practice.
\end{abstract}

\clearpage
\tableofcontents

\clearpage

\section{Introduction}
The basic forward problem of proton radiotherapy physics is finding the dose distribution in a known hetero\-geneous terrain exposed to known proton beams. We present a deterministic solution from first principles which differs in many respects from current pencil beam (PB) algorithms. We will use the acronym PBA (Pencil Beam Algorithm) for our algorithm and for the Fortran program written to test it.\footnote{~A pencil beam is just a bundle of protons. Its mathematical description is given in Sec.\,\ref{sec:PB}. We use the term `ur-beam' for an actual physical beam impinging on the terrain whereas `pencil beam' covers both ur-beams and the virtual beams generated in the calculation. Mathematically, they are identical.}

If the heterogeneity of the terrain is merely longitudinal, that is, if the terrain comprises successive uniform slabs of whatever material and thickness, a solution is already provided by Fermi-Eyges transport theory \cite{transport2012}. Then, however, all distributions up to and including the final dose are Gaussian, which fits few practical cases. 

In practice non-uniform slabs are present and a fundamental difficulty arises. Consider a brass collimator as a single slab, partly brass and partly air. Suppose a PB entering it overlaps both materials. Which one governs transport through the slab? Given the very different stopping and scattering powers of brass and air, that matters a great deal. The obvious choice is the material through which the PB centroid passes, the so-called `central axis' (CAX) approximation. However, that can lead to huge errors. If the CAX is just inside brass, the {\em entire} bundle of protons stops; if just outside, the {\em entire} bundle passes through with little scatter or energy loss.

The usual solution is to replace the ur-beam by an equivalent set of smaller PBs. That removes or mitigates the brass/air ambiguity for most of them. The dose at any point of interest (POI) in the terrain is the sum of doses from each of the new PBs, depending on its proximity to the POI. Now those PBs for which the CAX approximation is poor represent fewer protons, and the error is diluted by the other PBs.

\subsection{PB Redefinition and Dynamic Splitting}\label{sec:PBredefinition}
One method fitting the general description just given is PB {\em redefinition}, described some years ago by Shiu and Hogstrom \cite{shiu} (among others) in the context of electron beam dosimetry. At preselected intervals in $z$ (the beam direction) each PB is replaced by numerous smaller ones, while conserving the total number of incident particles and their phase space distribution. 

More recently, Kanematsu et al. \cite{kanematsuSplit} have proposed {\em dynamic splitting}. Here the daughter PBs are fewer and generated in a different manner, and the mother PB is split only when necessary, that is, when ambiguity of materials actually occurs. We will retain that `only when necessary' feature, but depart from \cite{kanematsuSplit} almost every other way, particularly recursion, which we permit and \cite{kanematsuSplit} suppresses.

Redefinition and dynamic splitting nominally address the same problem, but {\em we need both}. That essential point is best explained by looking ahead to our first example (Sec.\,\ref{sec:HCLdata}). A proton beam passes through a Pb scatterer and a large air gap and hits a brass collimator, where the single scattered Gaussian beam is purposely broad ($\sigma_x\approx80$\,mm) so as to illuminate the 20\,mm hole more or less uniformly. We wish to compute the dose downstream including collimator scatter (dose from protons that interact with the brass without stopping in it). 

That is dominated by a critical zone around the bore, $\approx1$\,mm thick for brass, where protons have a comparable chance of staying in the brass or scattering out (Fig.\,\ref{fig:GMCTR2}). Probing that zone requires PBs of comparable size, say $\sigma_x=0.5$\,mm, or $160\times$ smaller than the incident PB. As we will see, our splitting method reduces size by roughly $\sqrt{2}$ per split. It therefore requires some 15 splits. However, each split also replaces one PB by seven, yielding $7^{15}\approx5\times10^{12}$ PBs, an intractable number.\footnote{~We typically use about $10^{12}$ protons per treatment, so our smallest PB would represent a single proton!}

\enlargethispage*{1000pt}
Suppose instead that, just before the collimator, we fill a region a bit larger than the bore with overlapping 0.5\,mm PBs faithfully derived from the ur-beam, redefining it. That gets us within striking distance of the required size and generates only a few thousand PBs. Most will either stop in brass or pass through air without further division. In the example 425K PBs are generated in all. Of these, 117K cross a plane at the downstream face of the brass, combining into the six transverse dose distributions shown (Fig.\,\ref{fig:HCLX0258}). The calculation takes a mere 1.4\,min.\footnote{~All timings are for a single Lenovo T400 laptop, dual core 2.26\,GHz, 1.96\,GB RAM, running Compaq Visual Fortran under Windows XP Professional SP3.}
\pagebreak

We have explained why dynamic splitting alone (at least, as we implement it) is insufficient. On the other hand, redefinition alone generates many unneeded PBs, again yielding long computation times \cite{shiu}. Fig.\,\ref{fig:PBtracks}, generated by PBA, shows the combined effect of redefinition (collimator entrance) and subsequent recursive splitting. Note the qualitative resemblance to Fig.\,\ref{fig:GMCTR2}, though the assumed incident beam is less divergent. In both figures, straight-through tracks are suppressed.

\subsection{Other Features of PBA}
Besides combining redefinition with a recursive variant of dynamic splitting, PBA departs from current PB algorithms in other ways. 

We have already mentioned the {\em terrain} which comprises not just the patient or water tank but also (as desired) the scattering beam line and beam modifying devices such as range shifters, collimators, range compensators and air gaps. In other words, our algorithm solves (or can solve) the entire problem. Effective source distance and size are not required input. They emerge from the mechanical description of the beam line as we will demonstrate. 

Another feature is {\em parity of objects}. All items in the terrain are represented by stacks of slabs, heterogeneous or not as appropriate, and all such stacks are processed in exactly the same way. A collimator is simply a stack of heterogeneous slabs, thick enough to stop unlucky protons. There are no fictitious `beam defining planes' which either pass or totally stop protons. Fig.\,\ref{fig:SS2} illustrates how a compensated contoured scatterer, surrounded by its collimator, is handled by PBA. 

Yet another feature is {\em parity of materials}. Fig.\,\ref{fig:hiZloZ} shows that energy loss and multiple Coulomb scattering (MCS) behave quite differently as we traverse the periodic table. For water-like materials, one does sufficiently well by adjusting electron density, but for materials as disparate as Pb and Be we must use correct stopping and MCS theories. Therefore we do not single out any material (such as water) for special handling, or as a reference. Each material has its proper stopping power, using the  {\O}ver{\aa}s approximation to fit standard tables \cite{icru49,janni82}, and its proper scattering power using $T_\mathrm{dM}$ \cite{scatPower2010} as a proxy for Moli\`ere theory \cite{mcsbg}.

Our longitudinal or energy-like variable, rather than kinetic energy or residual range in water, is the kinematic quantity $pv$ (proton momentum times speed). $pv$ serves well for both the range-energy relation and MCS. Using it, rather than residual range in water, respects parity of materials.

The computational heart of PBA is a stack processor to handle splitting, nested in a loop over ur-beams, nested in a loop over states of the terrain (for instance, range modulator steps). That structure favors neither scattering nor pencil beam scanning (PBS). Either is handled with equal ease.

\subsection{A Test Program}
We developed and tested our algorithm with a Fortran program also called PBA. As usual at this stage of development, it treats simple geometric situations rather than being a dose engine for general use, but the code should be adaptable. Source code, a Windows\,XP executable and working files are in BGware at 
\begin{center}\BGsite\end{center}
We will use fragments of PBA or its input file to illustrate a number of points. A statement such as
\[\sigma_x\;:=\;[0.55]\;\sigma_x\]
means that $\sigma_x$ is replaced by its original value times a PBA parameter (usually from the input file) having a typical value of $0.55$. That is to say, $:=$ acts like $=$ in Fortran.

\section{Hard Scatters Excepted}\label{sec:hard}
Fig.\,\ref{fig:haloReactions} shows schematically the dose distribution around a beam stopping in water. It illustrates the fact that the {\em core}, dominated by multiple electromagnetic (EM) processes, is surrounded by a far larger {\em halo} (charged secondaries) and a still larger {\em aura} (neutral secondaries).\footnote{~Terminology from \cite{Gottschalk2015}, which see for details.} Those are due to hard single scatters either by the EM or the nuclear force, off free protons or the $^{16}$O nucleus as a whole (coherent scatter) or its constituent particles (incoherent scatter). 

At 180\,MeV, a typical radiotherapy energy, the core retains some 80\% of the primary protons and some 85\% of the primary energy (integrated dose). In common with all other PB algorithms, it is only the core whose evolution we calculate. That, however, dominates the high-dose region and therefore dominates the conformity of the dose to the target.

Pedroni et al. \cite{pedroniPencil} first showed that the halo must eventually be taken into account for accurate absolute dosimetry. How that is best done is a still unsolved problem in our view \cite{Gottschalk2015} and beyond the scope of this work. We shall say a bit more about it when discussing the appropriate mass stopping power $S/\rho$ to use when computing dose (Sec.\,\ref{sec:dose}).

\section{Collimator Scatter a Good Test}
Collimator scatter is a minor issue in proton radiotherapy, but it is the acid test of deterministic dose algorithms. The heterogeneity is extreme, and the zone of interest is $\approx1$\,mm depending on material. Moreover, experimental data are easily obtained with no more than a broad beam, a collimator, a small scanned dosimeter and perhaps a water tank. All this explains our seeming preoccupation with collimator scatter. 

\section{The Terrain}\label{sec:terrain}
Let us now describe the terrain in more detail. Everything is divided into slabs perpendicular to the nominal beam direction $\vec z/|z|$. The description corresponding to Fig.\,\ref{fig:HCLX0258} is
\begin{figure}[h]
\begin{verbatim}
terrain: mtl or file, #slabs, thickness (mm) - - - - - - - - - - - - - - -
LEAD          1       .5   material, # slabs, total mm
AIR           5   5853.    
REDEF1
HCL.CIR      40     36.5   file, # slabs, total mm, xbb min mm
HISTOGRAMS
AIR           1       .6   
AIR           4     40.    
AIR           1    134.8   
END         999    999     ================================================
\end{verbatim}
\end{figure}

\noindent The string `\texttt{terrain:}' identifies the block.\footnote{~Be careful about changing apparently harmless text in PBA.} The first line specifies 0.5\,mm thickness of lead (Pb) divided into 1 slab; the next, 5853.\,mm of air in 5 slabs. The next line is an instruction to transport the PB to the next slab, then redefine it. Parameters for the redefinition are elsewhere in the input file. The collimator itself is a bit too complicated to describe fully in the next line, so it refers to a file \texttt{HCL.CIR} (\texttt{.CIR} being our standard extension for a circular inhomogeneity) which reads
\begin{verbatim}
BRASS        outer material
AIR          inner material
0  0  9.88   xc,yc,radius (mm)
\end{verbatim}
\noindent Note, however, that the total thickness and number of slabs are always defined outside the file.\footnote{~That is handy for the simple geometries envisaged by PBA. A description of the patient would have to be much more complicated, but would still be sliced perpendicular to the nominal beam direction.}

The instruction `\texttt{HISTOGRAMS}' appears just prior to the $z$-plane at which histograms are to be incremented, and the list continues in the same vein. `\texttt{END}' designates the end of the terrain. A PB that has not ranged out is transported one last time to this plane, the dose is scored, and the PB is deleted.

A slab is identified with its entrance face. Thus $(pv)_5$ designates the value of $pv$ at the entrance to the fifth slab. Certain $z$-planes are designated as `measuring planes', at which the dose is evaluated. In PBA the measuring planes are contiguous and grow forward from \texttt{END} (they need not be equally spaced). In the example, if we specify (elsewhere) six measuring planes, the last is at $z_\mathrm{END}$ and the first at the entrance to the 40\,mm air gap.

To demonstrate that PBA is a calculation from first principles, suppose we wish to improve the penumbra at 175.4\,mm by replacing the 5.8\,m air gap by a helium bag. A conventional PB algorithm \cite{hong} would require new measurements of effective source distance and size. With PBA we need only replace \texttt{AIR} in the terrain description by \texttt{HELIUM} to obtain the equivalent of Fig.\,\ref{fig:HCLX0258} under the new conditions.

\section{Pencil Beam Parameters}\label{sec:PB}

There are eleven essential parameters plus two which are convenient in program development. PB parameters are only defined at a $z$ plane. The calculation starts by placing the parameters of an ur-beam at $z_1$ on a stack.\footnote{~The idea of using the stack is from Kanematsu et al. \cite{kanematsuSplit}. It transforms an indexing nightmare into a triviality.} In PBA the stack is implemented thus:
\begin{quote}\begin{verbatim}
 PARAMETER (k1=1,k4=20000)
 INTEGER*8 snbu
 COMMON/bunPars/k2,k3,snbu(k4),igbu(k4),xbu(k4),ybu(k4),jzbu(k4),
1  pvbu(k4),xpbu(k4),ypbu(k4),a0(k4),a1(k4),a2(k4),jzsp(k4),
2  bnc(k4)
\end{verbatim}\end{quote}
\texttt{k4} is the stack size which is large because thousands of PBs from redefinition may be placed on the stack before they are processed. \texttt{k3} is no longer used, \texttt{k1} is trivial, and \texttt{k2} is the stack pointer.\footnote{~PBA variable names often include the substring \texttt{bu} standing for `proton bundle'. `Pencil beam' implies smallness, and PBs can be very large, in single-scattered beams for instance. Nevertheless, in the body of this writeup we use the standard term.}

\subsection{Charge (2 Parameters)}
To maintain accuracy over a large number of binary splits we define an integral `generation index' (in the genealogical sense) \texttt{igbu\,}$=i_g=0,1,2,\ldots$. The charge carried by a given PB is proportional to $2^{-i_g}$. An ur-beam has $i_g=0$. When it is split one daughter has $i_g=2$ and the remaining six have $i_g=3$. Thus charge is conserved in a split ($1=2/8+6/8$). 

However, in implementing weighted beams (as in range modulation) or PB redefinition it becomes necessary to depart from binary arithmetic and we introduce a second charge-governing parameter \texttt{bnc\,}$\equiv q$ (`base nanoCoulombs'). The charge carried by a PB is therefore
\[\texttt{bnc(k2)}*(0.5**\,\texttt{igbu(k2))}\quad\mathrm{or}\quad q\times 0.5^{\;\displaystyle{i_g}}\quad\mathrm{nC}\]

\subsection{Position and Direction (5 Parameters)}
These are the usual $\texttt{xbu}=x,\,\texttt{ybu}=y$ (mm) and $\texttt{xpbu}=x'=\theta_x\,,\;\texttt{ypbu}=y'=\theta_y$ (mrad). 

The small-angle approximation is always valid for protons because the maximum MCS angle (for near-stopping Pb) is $\approx16$\degr\, and typical angles in a calculation are much smaller. However, because we slice the terrain into $z$-planes normal to the nominal beam direction, ur-beam inclinations are also limited by the small-angle approximation. For instance, if the gantry angle is changed significantly, that would call for a new calculation.

Because $z$ is quantized, the longitudinal coordinate is represented by an index \texttt{jzbu}.

\subsection{Energy-Like Variable (1 Parameter)}
We use the kinematic quantity $pv$ (proton momentum times speed) which is convenient for the range-energy relation (Section \ref{sec:RE}) and for multiple Coulomb scattering (Section \ref{sec:MCS}).\footnote{~Throughout this writeup we use $pv$ as a symbol for a {\em single} kinematic variable.} For a particle of rest energy $mc^2$ it is related to kinetic energy $T$ by
\begin{equation}\label{eqn:pv}
pv\;=\;\frac{\tau+2}{\tau+1}\;T
\end{equation}
where 
\begin{equation}\label{eqn:tau}
\tau\;\equiv\;T/mc^2
\end{equation}
is the reduced kinetic energy. For radiotherapy protons ($3\le T\le 300$\,MeV) the coefficient of $T$ in Eq.\,\ref{eqn:pv} ranges from 2 to 1.76 so one can think of $pv$ as being roughly twice the kinetic energy.

\subsection{Phase-Space Ellipse (3 Parameters)}
The Fermi-Eyges moments \cite{transport2012}
\begin{eqnarray}
A_0(j_z)&=&<\theta_x^2>\qquad\mathrm{(mrad^2)}\label{eqn:A0}\\
A_1(j_z)&=&<x\theta_x>\quad\mathrm{(mm\;mrad)}\label{eqn:A1}\\
A_2(j_z)&=&<x^2>\qquad\mathrm{(mm^2)}\label{eqn:A2}
\end{eqnarray}
represent the spread in projected angle and transverse position and the correlation between angle and position at the entrance to each slab. In a system of isotropic scatterers a PB has a circular cross section and its $A$s in the $x$ and $y$ projections are equal. In pencil beam scanning, incident beam characteristics may differ in the two transverse planes though we usually try to minimize that difference. If the algorithm were extended to magnets, we would definitely need two sets of As. In short, PBA assumes $A_\mathrm{n, x}=A_\mathrm{n, y}=A_\mathrm{n}$ but in some situations we might need two sets of $A$s.

PBA uses the $A$s internally but they are a bit abstract for I/O so we use proxies. Two choices are obvious: the gross transverse spread $\sigma_x=\sqrt{A_2}$ and the gross angular spread $\sigma_\theta=\sqrt{A_0}$. These define the bounding box of the ellipse (Fig.\,\ref{fig:ellipseThetaC}). 

A good candidate for the third is the angular confusion $\theta_c=B/A_2$ where $B\equiv A_0A_2-A_1^2$. It is an angle, which makes it easy to visualize, and it must lie between $\theta_c=0$ (a degenerate ellipse, emittance $=0$, position and direction perfectly correlated) and $\theta_c=\sigma_\theta$ (an erect ellipse, emittance $=\pi\sqrt{B}=\pi\,\sigma_x\,\sigma_\theta$). Physically, $\theta_C$ is the rms spread in direction of protons emerging from a slit at $x=0$ or any other $x$ \cite{transport2012}. It determines the sharpness of the dose distribution behind an edge illuminated by the PB in question.

The distance to the virtual point source $S_\mathrm{vir}=A_2/A_1$ also plays an important role in PBA, and one may wish to choose $\theta_c$ so as to obtain a desired $S_\mathrm{vir}$. Given $\sigma_x$, $\sigma_\theta$ and $\theta_c$ we find
\begin{equation}S_\mathrm{vir}\;=\;\sigma_x\times(\sigma_\theta^2-\theta_c^2)^{-1/2}\label{eqn:Svir}\end{equation}
so the desired value is
\begin{equation}\theta_c\;=\;(\sigma_\theta^2-(\sigma_x/S_\mathrm{vir})^2)^{1/2}\label{eqn:thetaC}\end{equation}
The desired $S_\mathrm{vir}$ can be arbitrarily large but it cannot be smaller than $\sigma_x/\sigma_\theta$ corresponding to the corner of the bounding box.

Eq.\,\ref{eqn:thetaC} illustrates a pitfall associated with expressing angles in mrad as PBA does (a choice which is otherwise convenient). We must convert $(\sigma_x/S_\mathrm{vir})$ to mrad before using it.

\subsection{Bookkeeping (2 Parameters)}
In program development it is convenient to assign a serial number \texttt{snbu} to each PB. Its largest value is then the total number of PBs generated in the calculation. (Some vanish and are never transported.) It may also be convenient to know where a given PB was created by setting a parameter \texttt{jzsp}
\,$\doteq j_z$ at split time and leaving it constant during transport.

\section{Redefinition}\label{sec:Redefinition}
Redefinition is the replacement of the mother PB by many daughters, much smaller and evenly spaced in the $xy$ plane. We usually use it at the entrance to a collimator and edit out daughters that are certain to stop. Four parameters govern redefinition. They are $p_1$ (mm) the $\sigma_x(=\sigma_y)$ of the daughters, $p_2$ (mm) their spacing, $p_3$ (mm) the margin allowed around the hole in a collimator and $p_4$ (dimensionless), the fraction of the mother to be used, in the scale of her $\sigma_x$. The last is needed when (for instance) the mother is much smaller than the hole in the collimator; thus $p_4=3$ means the mother PB will be `copied' out to $3\,\sigma$.

\subsection{Graphical Description}
Fig.\,\ref{fig:RedXY} shows daughters in the $xy$ plane at the entrance to a square collimator given a 1\,mm margin. Fig.\,\ref{fig:RedPS} shows mother and daughters in phase space. Each daughter is an erect ellipse with $\sigma_x=\sqrt{A_2}$ dictated by $p_1$ and height $\sqrt{A_0}$ equal to the angular confusion $\theta_\mathrm{C}$ of the mother. Since $\theta_\mathrm{C}$ is independent of $x$ that would, if we plotted the daughters in the normal way, give ellipses outside the mother ellipse, which looks wrong. Therefore we have instead adjusted the daughter ellipses for equal phase space density to the mother at their point of tangency.

\subsection{Algebraic Description}
Here and elsewhere in this document, square brackets identify parameters of the mother. Auxiliary quantities we will need are
\begin{eqnarray}
{[B\,]}&=&[A_0][A_2]-[A_1]^2\\
{[S_\mathrm{vir}]}&=&1000\; [A_2]/[A_1]\\
{[\,\theta_\mathrm{C}^2\,]}&=&[B]/[A_2]\\
r_{ij}^2&=&(x_i-[x])^2+(y_j-[y])^2\\
P(r_{ij}^2)&=&\frac{1}{2\,\pi\;\sigma_x^2}\;e^{\displaystyle{-\frac{1}{2}\frac{r_{ij}^2}{\sigma_x^2}}}\\
\noalign{\noindent Quantities governing spacing in $x$ and $x'$ (index $i$, columns) are\vspace{5pt}}
{[r_\mathrm{max}]}&=&p_4\;\sqrt{[A_2]}\\
\delta x&=&p_2\\
N_\mathrm{C}&=&\mathrm{MAX}\,(\,\mathrm{NINT}\,(2\,[r_\mathrm{max}]/\delta x),1)\\
\Delta x&=&(N_\mathrm{C}-1)\;\delta x\\
\Delta x'&=&1000\; \Delta x/[S_\mathrm{vir}]\\
f_i&=&-\,0.5+(i-1)/(N_\mathrm{C}-1)\quad,\quad i=1\ldots N_\mathrm{C}\\
\noalign{\noindent Quantities governing spacing in $y$ and $y'$ (index $j$, rows) are \vspace{5pt}}
\delta y&=&\sin(\pi/3)\;\delta x\\
N_\mathrm{R}&=&\mathrm{MAX}\,(\,\mathrm{NINT}\,(2\,[r_\mathrm{max}]/\delta y),1)\\
\Delta y&=&(N_\mathrm{R}-1)\;\delta y\\
\Delta y'&=&1000\; \Delta y/[S_\mathrm{vir}]\\
f_j&=&-\,0.5+(j-1)/(N_\mathrm{R}-1)\quad,\quad j=1\ldots N_\mathrm{R}\\
\noalign{\noindent Finally\vspace{3pt}}
q_{ij}&=&[q]\; 0.5^{\;\displaystyle{[i_g]}}\times F(\delta x)\times P(r_{ij}^2)\label{eqn:qij}\\
i_g&=&0\label{eqn:ig}\\
x_{ij}&=&[x]\;+\;f_i\,\Delta x\;+\;0.5\times\mathrm{MOD}\,(j+1,2)\times\delta x\\
x_{ij}'&=&[x']\;+\;f_i\,\Delta x'\;+\;0.5\times\mathrm{MOD}\,(j+1,2)\times\delta x'\\
y_j&=&[y]\;+\;f_j\,\Delta y\\
y_j'&=&[y']\;+\;f_j\,\Delta y'\\
\noalign{~}
j_z&=&[j_z]\\
pv&=&[pv]\\
A_0&=&[\theta_\mathrm{C}^2]\\
A_1&=&0\\
A_2&=&p_1^2
\end{eqnarray}

\noindent Eq.\,\ref{eqn:qij} apportions the electric charge of the mother (first factor on the RHS). $F(\delta x)$ corrects for the increase in the sum of daughters as they are brought closer together. (We tabulate the sum over a range of $\delta x$ values and fit that with a fourth degree polynomial, whose inverse is $F(\delta x)$.) $P(r_{ij})$ confers, on the sum of daughters, the cylindrical Gaussian shape of the mother. Charge is conserved to the extent that $p_4$ is sufficiently large. The generation index is cleared on redefinition (Eq.\,\ref{eqn:ig}) so that daughters get a fresh start vis-a-vis splitting.

\subsection{Gaussian Compliance}
Fig.\,\ref{fig:RedComp} demonstrates the Gaussian compliance of the sum of daughters to a mother partly passing through a centered collimator, with the mother offset in the $y$ direction.

\subsection{Trigger}
PBA triggers redefinition manually. One places `REDEF1' or `REDEF2' just before a collimator in the terrain description. Parameters for REDEF1 and REDEF2 may be the same or different.

\section{Splitting}\label{sec:Splitting}
Splitting a PB, in our sense of the term, replaces the mother PB by seven daughters. Except for the fact that our daughters, too, radiate from the virtual point source \cite{transport2012} defined by the mother, our scheme is entirely different from that described by Kanematsu et al. \cite{kanematsuSplit}. We attempt to preserve, as well as possible, the cylindrical symmetry characteristic of MCS in ordinary materials. That is, our scheme is inherently 2D. Moreover, we conserve emittance $\pi\sqrt{B}$ rather than the local angular spread or `angular confusion' $\theta_\mathrm{C}$.

\subsection{Graphical Description}
Fig.\,\ref{fig:SplitXY} shows mother and daughters in the $xy$ plane at the $z$-plane where the split occurs. Circle radii are proportional to $\sigma_x\,(=\sigma_y)$ of each PB. Two dimensionless parameters govern the split: $p_1\approx0.55$ the change in $A_0$ and $A_2$, and $p_2\approx1.03$ the spread in $xy$ space. Thus daughters have $\sigma_x=\sqrt{p_1}\;[\sigma_x]$ and $\sigma_x$ only decreases by $\approx3/4$ per split. Trying for more worsens the Gaussian compliance discussed later. 

Line weights in Fig.\,\ref{fig:SplitXY} are proportional to the relative charge carried by each PB: 8 units for the mother, 2 for the central daughter and 1 for the six outboard daughters corresponding to generation indices $i_g=0,\,2,\,3$. Thus splitting skips a generation, and charge is rigorously conserved in a split.

Fig.\,\ref{fig:SplitPS} shows the split in phase space, with four daughters that would otherwise overlap slightly displaced.

\subsection{Algebraic Description}\label{sec:seven}
Vector notation is convenient. Defining auxiliary unit, generation ($i_g$), cosine and sine vectors by
\begin{eqnarray}
\mathbf{U}&\equiv&(1,1,1,1,1,1,1)\\
\mathbf{G}&\equiv&(2,3,3,3,3,3,3)\\
\mathbf{C}&\equiv&(0,\cos(0),\cos(\pi/3),\cos(2\pi/3),\cos(\pi),\cos(4\pi/3),\cos(5\pi/3))\\
\mathbf{S}&\equiv&(0,\sin(0),\sin(\pi/3),\sin(2\pi/3),\sin(\pi),\sin(4\pi/3),\sin(5\pi/3))
\end{eqnarray}
and, as usual \cite{transport2012}
\begin{eqnarray}
[S_\mathrm{vir}]&=&[A_2]/[A_1]\\
{[B]}&\equiv&[A_0]\,[A_2]-[A_1]^2\\
\noalign{\vspace{3pt}\noindent then, with  daughters indexed per Fig.\,\ref{fig:SplitXY}, we have\vspace{5pt}}
\mathbf{i_g}&=&[i_g]\,\mathbf{U}\;+\;\mathbf{G}\\
\mathbf{q}&=&[q]\;\mathbf{U}\\
\mathbf{x}&=&[x]\,\mathbf{U}\;+\;p_2\,[\sigma_x]\,\mathbf{C}\\
\mathbf{x'}&=&[x']\,\mathbf{U}\;+\;(p_2\;[\sigma_x]\,/[S_\mathrm{vir}])\,\mathbf{C}\\
\mathbf{y}&=&[y]\,\mathbf{U}\;+\;p_2\,[\sigma_y]\,\mathbf{S}\\
\mathbf{y'}&=&[y']\,\mathbf{U}\;+\;(p_2\,[\sigma_y]\,/[S_\mathrm{vir}])\;\mathbf{S}\\
\mathbf{j_z}&=&[j_z]\;\mathbf{U}\\
\mathbf{pv}&=&[pv]\;\mathbf{U}\\
\mathbf{A_0}&=&p_1\,[A_0]\;\mathbf{U}\\
\mathbf{A_1}&=&(\,p_1^2\;[A_0]\,[A_2]\,-\,(1/7)^2\;[B]\,)^{1/2}\;\;\mathbf{U}\\
\mathbf{A_2}&=&p_1\,[A_2]\;\mathbf{U}
\end{eqnarray}

\subsection{Gaussian Compliance}
How well does the sum of split daughter PBs match the original Gaussian? We apply an ur-beam to MP $z_1$, which automatically records the dose distribution there. We then split the PB recursively into 751 daughters (arbitrary number), transport them through 0.1\,mm air, and record their summed dose distribution at $z_2$. The two dose distributions agree closely along the $x$ axis (Fig.\,\ref{fig:SplitX}) as well as the $y$ axis (not shown).

Next, we directed a 158.6\,MeV ur-beam at a water tank and transported it unsplit to 50, 100 and 150\,mm (near stopping) depth. In a second run we split the ur-beam into 1219 daughters (arbitrary number) before transporting them. In this case, transport involves transverse spreading in the water as well as the increase of stopping power with depth. Again, the two dose distributions agree in $x$ (Fig.\,\ref{fig:HMLUtest}) and $y$ (not shown). 

Agreement in both $x$ and $y$ is non trivial since they correspond to different planes of symmetry cf. Fig.\,\ref{fig:SplitXY}. We have verified, in numerous cases where cylindrical symmetry is expected, that dose distributions along an axis inclined at an arbitrary angle agree closely with those along the $x$ axis.

\subsection{Trigger}\label{sec:splitTrigger}
To split or not to split is governed by three parameters $p_1\approx2.4$, $p_2\approx0.6$ and $p_3\approx7$. Their numerical values vary widely depending on the problem and the degree of convergence required. Each PB is examined at each $z$-plane and split if 
\begin{equation}\label{eqn:splitTrigger}
i_g\le p_3\quad\mathrm{AND}\quad A_2>p_2\quad\mathrm{AND}\quad d<p_1\,\sigma_x
\end{equation}
The first condition tests the generation index, limiting the depth of recursion. Depending on the problem, execution time depends strongly on $p_3$, often doubling or worse per unit increase in $p_3$. The second condition requires the PB to have some fixed size before it splits. The third is the most obvious, $d$ being the distance to a boundary between two materials. It requires the PB axis to be near a boundary, in the scale of the PB's $\sigma_x$. We do not, at present, test the {\em degree} of heterogeneity, that is, how different the two materials are. Virtual heterogeneities such as an air/air boundary are permitted, and are useful for such tests as described in the previous section.

At present the splitting trigger Eq.\,\ref{eqn:splitTrigger} is the least satisfactory aspect of the algorithm. Although it is possible to find combinations of $p_{1,2,3}$ that give solutions reasonably stable with respect to these and other parameters, we do not understand the interaction between $p_{1,2,3}$ very well, and have so far been unable to find one single parameter that governs convergence smoothly. Furthermore, though some absolute condition such as $A_2>p_2$ appears to be necessary, it introduces non-physical discontinuities into the $pv$ distribution and sometimes into final dose distributions. More work on the trigger is needed.

\section{Transport}
To `transport' a PB from $z_i$ to $z_{i+1}=z_i+\Delta z$ is to replace its parameters at $z_i$ by those at $z_{i+1}$. The equations, where `$:=$' reads `is replaced by' (same as Fortran `='), are
\begin{eqnarray}
i_g&:=&i_g\\
q&:=&q\\
x&:=&x\;+\;x'\,\Delta z\\
x'&:=&x'\\
y&:=&y\;+\;y'\,\Delta z\\
y'&:=&y'\\
j_z&:=&j_z\;+\;1\\
pv&:=&PV\,\left(R\,(\,pv\,,M)\;-\;\Delta z\right)\label{eqn:RE}\\
A_2&:=&A_2\;+\;(2A_1+(A_0+(\tilde{T}/3)\Delta z)\Delta z)\Delta z\label{eqn:a2k}\\
A_1&:=&A_1\;+\;(A_0+(\tilde{T}/2)\Delta z)\Delta z\label{eqn:a1k}\\
A_0&:=&A_0\;+\;\tilde{T}\Delta z\label{eqn:a0k}
\end{eqnarray}
For instance, Eq.\,\ref{eqn:RE} tells us to find the range corresponding to $pv$ in material $M$, subtract $\Delta z$, the step thickness, to get the residual range, convert back to $pv$ and replace the current $pv$ by that smaller value.
Pursuant to these formulas charge is conserved, $x$ and $y$ propagate as straight lines, slopes are conserved, and the $z$-plane index increases by 1. Material properties enter through the range-energy relation (transport of $pv$) and the $A$\,s (MCS).

\subsection{Range-Energy Relation}\label{sec:RE}
The new value of $pv$ is computed by manipulating the range-energy relation which (unlike using stopping power) is correct for any $\Delta z$. We use the {\O}ver{\aa}s approximation \cite{overas}
\begin{equation}\label{eqn:Overas}
(pv)^2\;\approx\;aR\,^b
\end{equation}
where $R$ is the geometric range (mm) and $a,b$ are parameters characteristic of the stopping material. At initialization PBA assigns an index to each new material and finds its $a$ and $b$ using two kinetic energies (input parameters) and the corresponding ranges from standard tables e.g. \cite{icru49,janni82}.

Fig.\,\ref{fig:TestRover} compares Eq.\,\ref{eqn:Overas} to the more common $R=aT^b$. They have comparable accuracy (1-2\%) but are seen to be complementary, {\O}ver{\aa}s being better for light materials and $R=aT^b$ for heavy materials. Details depend on the fiducial energies (here 32 and 160\,MeV), and it is possible to make one approximation look considerably better than the other in particular cases. 

\subsection{Scattering Power}\label{sec:MCS}
Eqs.\,\ref{eqn:a2k} through \ref{eqn:a0k}, from Kanematsu \cite{kanematsu08}, are equivalent to standard transport equations for the $A$\,s \cite{transport2012} integrated by the midpoint method. $\tilde{T}$ is the value at mid-step of the scattering power
\begin{equation}\label{eqn:Txx}
T_\mathrm{xx}(z)\;\equiv\;\frac{d<\theta_x^2>}{dz}
\end{equation}
(expressed in mrad$^2$/mm) for which a number of formulas have been proposed \cite{scatPower2010}.\footnote{~In this paper, $T$ with letter subscripts denotes a scattering power rather than kinetic energy.} We use
\begin{equation}\label{eqn:TdM}
T_\mathrm{dM}\;\equiv\;f_\mathrm{dM}((pv)_1,pv)\times\left(\frac{E_s}{pv}\right)^2\;\frac{1}{X_S}
\end{equation}
where $E_s=15.0$\;Mev. The correction factor, which measures nonlocality by the decrease in $pv$, is
\begin{eqnarray}\label{eqn:fdM}
f_\mathrm{dM}&\equiv&0.5244+0.1975\log_{10}(1-(pv/(pv)_1)^2)+0.2320\log_{10}(pv/\mathrm{MeV})\nonumber\\
&&-\;0.0098\log_{10}(pv/\mathrm{MeV})\log_{10}(1-(pv/(pv)_1)^2)
\end{eqnarray}
$(pv)_1$ is the initial value and $pv$ is the value at the point of interest. The scattering length $X_S$ (cm) is 
\begin{equation}\label{eqn:XS}
\frac{1}{\rho X_S}\;\equiv\;\alpha N r_e^2\,\frac{Z^2}{A}\left\{2\log(33219\,(AZ)^{-1/3})-1\right\}
\end{equation}
where $\rho$ is density, $\alpha$ is the fine structure constant, $N$ is Avogadro's number, $r_e$ is the classical electron radius and $A,Z$ are the atomic weight and atomic number of the scattering material. For compounds and mixtures $X_S$ obeys a Bragg rule 
\begin{equation}\label{eqn:XSBragg}
\frac{1}{\rho X_S}\;=\;\sum_iw_i\left(\frac{1}{\rho X_S}\right)_i
\end{equation}
where $w_i$ is the fraction by weight of the $i^\mathrm{th}$ constituent. $X_S$ is a material property very similar to the radiation length $X_0$ but it improves the material dependence of the scattering power particularly for light materials. For details see \cite{scatPower2010} and for a table of $X_S$ values for common materials see \cite{transport2012}. Figs.\,\ref{fig:ExptFigPoly} and \ref{fig:ExptFigLead} compare $T_\mathrm{dM}$ for polystyrene (nearly water equivalent) and lead with theory, with experiment, and with other formulas $T_\mathrm{xx}$.

\subsection{RangeOut and ReachEnd}
If the proton ranges out in $\Delta z$ (residual range $\le0$) the transport subroutine sets a logical variable \texttt{RangeOut} which will cause the main program to go to the next PB. If, alternatively, $z_{i+1}$ corresponds to the last plane in the problem, \texttt{ReachEnd} is set with the same result.

\section{Basic Program}
Fig.\,\ref{fig:basicProgram} is a heavily edited version of the basic program, which consists of three nested loops. The outermost runs over states of the terrain (here exemplified by \texttt{np1} and \texttt{np2}, the indices of the first and last range modulator steps). Since every state of the terrain is different and may be exposed to different ur-beams, \texttt{WhatsHere} (explained later) and the array of ur-beams are initialized.

The next loop pushes the state of the next ur-beam at $z_1$ onto the computer stack and scores its dose at $z_1$ if $z_1$ is a measuring plane.

In the innermost loop, the state of the current PB is examined at the current $z$-plane by a subroutine \texttt{WhatsHere}. Given $x$ and $y$ of the PB centroid as well as the $z$-plane index $j_z$, \texttt{WhatsHere} returns \texttt{kml}, the material index at that position on that $z$-plane, \texttt{dzz}, the thickness of the next slab, the density and scattering length of the material and above all, \texttt{dsqnh}, the squared distance to the nearest heterogeneity boundary. (In a homogeneous slab \texttt{dsqnh} is set to some very large number.)

Depending on \texttt{WhatsHere} one of three things happens. The PB state is
\begin{itemize}
\item Redefined (replaced by many new PBs) or
\item Split (replaced by seven new PBs) or
\item Transported through the next slab (PB state including \texttt{jz} is replaced).
\end{itemize}
as detailed in previous sections.
If during transport a PB ranges out it is deleted (the stack pointer \texttt{k2} is decremented). If it reaches the last plane in the problem its dose is scored there and the PB is deleted. When the stack is empty (\texttt{k2 = 0}) control passes out to the ur-beam loop which places the state of the next ur-beam (at $z_1$) onto the stack. When the ur-beams for the current state of the terrain are exhausted control passes to the terrain loop and the next state of the terrain. When all states of the terrain are exhausted the computation is done.

So far, no useful information has been saved from all this activity. After each exit from \texttt{Transport}, if the $z$-plane just reached is a measuring plane, \texttt{AddToDose} updates a dose matrix by adding to appropriate elements the dose from the PB just transported. The dose matrix (defined to suit the problem under study) is the permanent record of what was accomplished and, along with the input file (a record of what was undertaken) is available to PBA and other programs for post-processing (`analysis').

Note that the program makes no distinction between scattered and scanned (PBS) beams. Ins scattered beams the terrain will be more complicated and usually have more states (range modulator steps), while there will only be one ur-beam per step. In PBS the reverse is true, but the program is the same in either case. 

\section{Fluence and Dose}\label{sec:dose}
The program described thus far is basically a fluence calculation. At the current measuring plane we know the centroid $x,y$ and rms width $\sigma_x$ of a cylindrical Gaussian describing the fluence of the current  PB, as well as its energy-like variable $pv$. We score the dose deposited by that PB at selected measuring points in the plane\footnote{~In PBA those points are limited to two axes, usually the $x$ and $y$ axes, perpendicular to the $z$ axis.} using the fundamental equation
\begin{equation}\label{eqn:D}
D\;=\;\Phi\,\times\,S/\rho
\end{equation} 
where $D$ is dose, $\Phi$ is fluence and we will deduce the mass stopping power $S/\rho$ from the value of $pv$.

Here a problem mentioned in Sec.\,\ref{sec:hard} arises, namely the dose from hard single scatters (or `nuclear interactions'). That dose forms a broad halo around the EM core (cf. Fig.\,\ref{fig:haloReactions}). Accounting for it properly in arbitrary geometries is complicated, and is an open question. Here, we will assume one of two limiting cases:
\begin{itemize}
\item The POI is in the middle of a broad beam where EM and `nuclear' equilibrium obtain, or
\item The POI is in a region where there are few secondaries from hard scatters.
\end{itemize}
In the first case an effective stopping power $S_\mathrm{mixed}$, mixing EM dose with nuclear secondary dose, is appropriate; in the second, a purely EM stopping power $S_\mathrm{em}$ \cite{Gottschalk2015,Gottschalk2014}.

\subsection{Mixed Stopping Power}
$S_\mathrm{mixed}$ is appropriate near the middle of a broad field in (say) a water tank. It cannot be calculated from first principles because there is no convenient theory describing hard scatters. It can, however, be measured by creating a broad beam, measuring the depth-dose distribution along its central axis, and dividing the dose by the fluence in air at the time of measurement, applying Eq.\,\ref{eqn:D}.\footnote{~This is in essence a $1/r^2$ correction. The effective source distance must be known.} 

In PBA the depth-dose is retrieved from a text file, extension .BPK, containing a cubic spline fit to the measured data. At initialization this file is read, corrected for $1/r^2$, normalized to absolute Gray/($10^9$p/cm$^2$)) using the tabulated $S/\rho$ at the incident energy, and used to populate a linear interpolation lookup table.

In current practice ($S_\mathrm{mixed}/\rho_\mathrm{water}$) is known as integral depth dose (IDD) and is used (wrongly, we believe) to parameterize the core/halo and commission PBS treatment planning systems \cite{Gottschalk2015,pedroniPencil}. 

\subsection{EM Stopping Power}
$S_\mathrm{em}$, unlike $S_\mathrm{mixed}$, cannot be measured because we cannot turn off hard scatters. However, it can be computed from first principles by convolving a tabulated $-(dT/dz)$ with a Gaussian of rms width $\sigma_\mathrm{sem}$ which combines range straggling with the energy spread (if any) of the incident beam. Thus 
\begin{equation}\label{eqn:convo}
S_\mathrm{em}(z)\,=\,\int_{z-5\sigma}^{z+5\sigma}-\frac{dT}{dz}(z)\;G(z-z',\sigma_\mathrm{sem})\,dz'
\end{equation}
where $G$ is a 1D Gaussian normalized to unit area. The limits of integration are arbitrary and are adjustable in PBA. Integration may be performed numerically by Simpson's Rule \cite{nr}, but requires a cutoff procedure because $-dT/dz$ is singular at end-of-range. We break the integral at $z=z_c$\, into two terms, one nonsingular and the other susceptible to approximation: 
\begin{equation}\label{eqn:cutoff}
S_\mathrm{em}(z)\,=\,\int_{z-5\sigma}^{z_c}-\frac{dT}{dz}(z)\;G(z-z',\sigma_\mathrm{sem})\,dz'\,+\,
  \int_{z_c}^{z+5\sigma}-\frac{dT}{dz}(z)\;G(z-z',\sigma_\mathrm{sem})\,dz'
\end{equation}
The second integrand is dominated by the singularity of $-dT/dz$ at $R_0\approx z_c$. We approximate that integral by setting $z'=z_c$ and find
\begin{equation}\label{eqn:Sem}
S_\mathrm{em}(z)\,\approx\,\int_{z-5\sigma}^{z_c}-\frac{dT}{dz}(z)\;G(z-z',\sigma_\mathrm{sem})\,dz'\,+\,
  T(R_0-z_c)\,G(z-z_c,\sigma_\mathrm{sem})
\end{equation}
$T(R_0-z_c)$ is the residual kinetic energy, which we obtain from a range-energy table, at $z_c$. The overall result is insensitive to $z_c$.

\subsection{Energy Extension}
A minor problem arises in that the mean projected range of a proton beam equals the distal 80\% point of the Bragg peak or effective stopping power: $R_1=d_{80}$.\footnote{~This important relation, first proved by Andy Koehler, has been confirmed many times since then cf. \cite{bortfeld}} If range-energy tables are used the normal way, the calculation terminates (\texttt{RangeOut} is set) at $d_{80}$. Sometimes is is desirable to see the dose distribution beyond that point. In PBA that is accomplished by setting the `BP toe level' to (say) 0.2 rather than 0.8. Then incident $T$ (and therefore incident $pv$) are increased slightly to postpone \texttt{RangeOut}. That results, of course, in a slight error, but we have not thought of a better workaround. The problem may stem from approximating, by a deterministic algorithm, what is fundamentally a stochastic process (slowing down of protons).

\subsection{Dose Accumulation}
Using Eq.\,\ref{eqn:D} we add the appropriate dose to all measuring points in the current measuring plane within range of the current PB. We begin with the point nearest the PB centroid and work outwards in each direction until the fluence vanishes. The fluence is given by a 2D (cylindrical) Gaussian of rms width $\sigma_x$ centered at $(x,y)$, the coordinates of the centroid. The appropriate effective mass stopping power in the `dose-to' material, either measured or computed as just described, has previously been tabulated as a function of residual range. (For instance, if we are concerned with diode measurements, the `dose-to' material would be silicon.) The residual range is computed from $pv$ using the {\O}ver{\aa}s approximation.  

Obviously we can accumulate fluence (instead of dose) distributions by setting the stopping power to an appropriate constant. A PBA software switch selects fluence, $(\mathrm{Mp/cm}^2)/\mathrm{(incident\ nC)}$, instead of dose, $\mathrm{mGy}/\mathrm{(incident\ nC)}$ .

\section{Example: a Harvard Cyclotron Experiment}\label{sec:HCLdata}
Around 1993, while investigating off-energy protons in dosimeter calibration beams \cite{verhey}, we measured collimator scatter at the Harvard Cyclotron Laboratory (HCL). These data, presented at conferences \cite{Mayo1993,bgslit} but never published, were invaluable in developing PBA, and we now discuss them at considerable length. Unfortunately, original tapes and notes are long gone. We re-constituted the data from old graphs, which may partly account for the problem noted below.  

We will discuss (1) the experiment, (2) the systematics of collimator scatter, (3) a Monte-Carlo (TOPAS) study of hard scatters, (4) where to normalize PBA to the experimental data, (5) the $1/r^2$ dependence of fluence, (6) PBA convergence, (7) pristine and degraded protons studied separately and (8) anomalies in certain frequency distributions. Finally (9) we will summarize what we have learned and what needs to be re-done.

We use the notation $D_\mathrm{MP}(r)$; thus $D_1(\approx0)$ denotes the dose at measuring plane 1 (0.6\,mm from the downstream collimator face) at a radius near 0\,mm. Protons reaching MP1 with only the nominal energy loss to Pb and air in the beamline are `pristine'. Those that, in addition, lose energy to brass are `degraded'. The former have $pv>284$\,MeV; the latter, $0<pv<284$\,MeV.\pagebreak 

\subsection{The Experiment}
A single scattered 158.6\,MeV proton beam was directed at a brass collimator having a cylindrical bore of 9.88\,mm radius. (For exact dimensions see Sec.\,\ref{sec:terrain}.) Full transverse scans in air were taken with a small diode at six distances (MP\,1\,-\,6) from the downstream face of the collimator. Right and left half-scans were averaged to yield six virtual radial scans. The experimental data are shown, for instance, in Fig.\,\ref{fig:HCLX0258}, along with a 1.4\,min PBA computation (line) already mentioned (Sec.\,\ref{sec:PBredefinition}). 

A single constant, derived from $D_1(\approx0)$ and shown in all panels for reference, normalizes the PBA results. The rms deviation of PBA from experiment hovers around 6\,-\,7\% for all MPs. The choice of $D_1(\approx0)$ rather than  $D_6(\approx0)$ is deliberate and will be discussed. The latter choice yields a far better fit, $\approx1.4\%$, to MP6. 
 
\subsection{Systematics of Collimator Scatter}
The systematics of collimator scatter \cite{courant,BGcourse} can be inferred from Fig.\,\ref{fig:GMCTR2}. Protons that interact with brass but survive fall into two categories. Those entering just outside the bore (`outers') converge towards the axis with broad angle and energy distributions: for brass, the mean angle is $\approx6$\,\degr\ and the critical zone around the bore is $\approx1$\,mm \cite{BGcourse,burge}. Those entering via the bore either scatter back out or emerge from the downstream face with a broad distribution of angles peaked outwards and energies peaked at the high end, from protons grazing the distal corner. These `inners' are, however, scarce in our data because the incident beam was nearly parallel.

Thus, near the collimator, we see a sharp spike just inside the physical radius: protons heading towards the axis but still far from it. Dose on axis near the collimator $D_1(\approx0)$ is almost all from pristines, if the dose from hard scatters is small. Moving downstream the spike broadens until, far from the collimator at MP6, most degraded protons have crossed the axis and are thinly spread, though not negligible as we will see.

\subsection{A Monte Carlo (TOPAS) Study}
Since PBA ignores hard scatters (Sec.\,\ref{sec:hard}) it was important to estimate their contribution to the experiment. David Hall of the Massachusetts General Hospital (MGH) kindly ran two Monte Carlo TOPAS simulations, the first including all processes, the second omitting those involving the nuclear force. Appendix \ref{sec:TOPAS} gives details. Fig.\,\ref{fig:TOPAS} shows the experimental data along with the TOPAS `all' and `EM only' results normalized in this case to $D_6(\approx0)$, using a single constant for both runs and all MPs. General agreement is excellent, bolstering our confidence in both TOPAS and the experiment. The `all' and `EM only' runs are barely distinguishable, justifying the view that hard scatters play a minor role.

\subsection{Choosing a Normalization Point}
The choice of $D_1(\approx0)$ vs. $D_6(\approx0)$ as the normalization point is more important than might appear. In normalizing one set of curves to another one usually seeks to give the best overall impression. Here that favors $D_6(\approx0)$ (cf. Fig.\,\ref{fig:HCLX0258}) which would yield excellent agreement at MP5\,-\,6 but a PBA shortfall of some 6\% at $D_1(\approx0)$. The conundrum is this: the obvious choice suggests a PBA shortfall on axis near the collimator, the other a PBA excess distant from the collimator. In trouble-shooting PBA, it matters a great deal which of the two we are looking for. Let us take a closer look at what we expect at MP6 relative to MP1. We expect a diminution of dose from $1/r^2$ but an increase in dose from degraded protons, spread out but still there. 

\subsection{$1/r^2$ Dependence of Fluence}
The virtual point source \cite{transport2012} of a Gaussian beam is used by PBA during redefinition (Sec.\,\ref{sec:Redefinition}). We choose the point source for simplicity if nothing else: it would be difficult to populate an extended source in a deterministic algorithm. Now the virtual point and effective extended sources \cite{transport2012} are at different distances $S$ from MP1. Under our conditions $S_\mathrm{vir}\approx5800$\,mm and $S_\mathrm{eff}\approx5400$\,mm, whereas the physical distance from the Pb foil to MP1 is 5890\,mm. Could this somehow be responsible for the 6\% tension between $D_1(\approx0)$ and $D_6(\approx0)$?

Put more precisely, how much does the actual fluence $\Phi(z)$ on axis differ from the $1/r^2$ dependence of a point source? Analysis (Appendix \ref{sec:inverseRsq}, Fig.\,\ref{fig:fluenceCorr}) shows the effect is small and constant over the applicable range of $z$. Therefore this explanation of the inconsistency can be ruled out.

\subsection{Analysis of Pristine and Degraded Protons}
A simple cut at $pv=284$\,MeV separates pristine from degraded protons allowing us to study them separately. Fig.\,\ref{fig:HCLX0261} is a normal run with dose accumulated only for pristine protons. The PBA results are normalized at $D_1({\approx0})$ and the defect at $D_6(\approx0)$ is exactly that expected from $1/r^2$. Fig.\,\ref{fig:HCL0264} displays degraded protons at the six measuring planes.  The data points are absolute dose from a home-made Monte Carlo GMC. They (and the PBA lines) show a rather considerable contribution from degraded protons at MP6, though quite spread out by then as expected.

Incidentally, this is also a good test of the cylindrical symmetry expected in this experiment. There is a considerable difference between the dose on the $x$ and $y$ axes at MP5 at the geometric radius of the hole. Elsewhere the agreement is fair.

\subsection{Convergence}
The HCL data afford us a simple way of studying PBA convergence. On general principles we expect that reproducing very sharp features in the dose distribution requires superposing very small PBs, with a steep increase in computation time. That is confirmed by Figs.\,\ref{fig:HCLX0258} and \ref{fig:HCLX0262}. The former corresponds 0.53\,M PBs traversing the region of interest MP1~-~MP6; it took 1.4\,min. The latter (recursion cutoff raised from 10 to 18) corresponds to 40\,M PBs $(75\times)$ in the ROI, and took 93\,min $(66\times)$. Compliance at MP1 indeed improved from 7.8\% to 6\% but elsewhere it was worse and overall, the same. Thus, execution time depends exponentially on the sharpest features to be reproduced, and broader features may actually suffer. The submillimeter edge at MP1 is a very stringent test.

\subsection{Frequency Distributions of PBA Quantities}
PBA includes a histogrammer to study frequency distributions of selected variables. Semilog Fig.\,\ref{fig:Logh0262} corresponds to the Fig.\,\ref{fig:HCLX0262} run with histograms populated at MP6. Variables shown (CW from upper left) are $pv$, $pv$ with coarser bins, $\sigma_x$ (\texttt{xbb}), $\sigma_\theta$ (\texttt{ybb}), \texttt{jzsp} the $z$-plane index where the PB was born, and $i_g$ the generation index ($\log_{\,2}$ of the charge). For instance, at MP6 the most likely PB size is $\approx0.5$\,mm.

Non-physical spikes are evident in the continuous variables $pv$, $\sigma_x$ and $\sigma_\theta$, apparently a beat effect between the absolute size cut in the splitting trigger (Sec.\,\ref{sec:splitTrigger}) and the number of slabs representing the collimator. This is not as serious as might appear from the frequency distribution, the absolute $pv$ difference between `popular' and `normal' values being small. Nevertheless we may be seeing traces of it in a depth-dose studied later (cf. Fig.\,\ref{fig:Hong11}). This pathology of PBA may be another consequence of computing deterministically a process which is fundamentally stochastic, dose deposition by protons.

\subsection{Discussion}
We have discussed the HCL data at length, dwelling on the discrepancy between experiment and PBA at MP6 where the agreement should be good (and is, if we normalize there). That preoccupation should not detract from the fact that PBA computes most of the key features of the dose, quickly solving from first principles a problem that no other deterministic algorithm even attempts. 

We focused on the MP6 anomaly, and where to normalize, because that determines whether we look for a PBA excess in one place or a defect in another, a distinction critical to trouble-shooting. In the end, we reject both hypotheses, concluding there is a problem with the experimental data. The distribution at MP6 should be some 6\% higher and the tail beyond the edge, from degraded proton crossovers, should be more pronounced. Whether the problem is due to reconstructing data from graphs, a faulty drift correction at the time of the experiment, or some other cause, is impossible to say. Ideally this simple experiment, or one very like it, should be repeated. 

\section{Example: Half-Beam Block at 218\,MeV}

\subsection{Background}
In 2006 Slopsema and Kooy \cite{slopsema} and independently, Kanematsu et al. \cite{kanematsu06} studied and parameterized the effect of finite collimator thickness. They model a half-beam block as a pair of ideal absorbing planes separated by the actual thickness. In a scattered beam line, the angular confusion is the same on- and off-axis \cite{transport2012} but the mean proton angle is not. In consequence the two fictitious on-axis edges intercept roughly equal numbers of proton trajectories, whereas off-axis the downstream edge intercepts far more than the upstream edge, defining the spatial distribution more cleanly. Therefore the on-axis penumbra of a half-beam block in a double scattered beam was predicted, and confirmed by related measurements, to be slightly greater than the off-axis penumbra.

In 2002 Kooy, while commissioning Gantry\,1 at the MGH Burr Center \cite{kooy}, had obtained data also relevant to this problem. To determine effective source distances and sizes for each of the seven range options, and for zero and full range modulation, he took transverse scans in air with a PTW 31006 \cite{ptw} thimble chamber (inner diameter 2\,mm) at 200 and 700\,mm from the downstream face of a $40\times200$\,mm half-beam block: 28 scans in all. Here we will consider just one: option A7 (range 22\,cm H$_2$O), no range modulation, 200\,mm air gap. 

\subsection{PBA Setup and Computation}
Fig.\,\ref{fig:HBB0280} shows the measurement and the PBA prediction while Fig.\,\ref{fig:HBBfiles} shows the entire PBA input file, including the terrain.\footnote{~We replaced three Pb prescatterers, used for dose flatness correction, by one of the same total thickness.} The computation took 14.6\,min. 86\,M PBs were generated of which 46\,M reached the measuring plane, combining to form the dose distribution shown. Since the PBs carry very different charges, a better throughput measure is that, of 1043\,pC, 100 reach the MP while 943 range out in the  collimator around the contoured scatterer or in the half-beam block itself.\footnote{~The incident charge is exactly 1000\,pC but redefinition introduces some error. PBs edited out during redefinition are counted as ranged out.}

We start with an ideal monoenergetic beam because the scatterers in the beam line swamp any reasonable initial spread, divergence and emittance, and the machine energy spread is negligible compared with $\sigma_\mathrm{R}/R\approx1.1\,\%$ expected \cite{janni82} from range straggling and used in computing $S_\mathrm{em}$. 

\subsection{Results}
The on-axis and off-axis (L and R) $80/20$ penumbras (Fig.\,\ref{fig:HBB0280}) are 5.11, 4.65\,mm (PBA) and 5.26, 4.86\,mm (measured), corresponding to -0.15\,mm and -0.21\,mm, or -2.8\% and -4.3\%, deviations of PBA from measurement. Some of this is attributable to detector size, correction for which would lower the measured values by roughly $\sqrt{5^2-1^2}/5=0.98$. The excess of the on-axis over the off-axis penumbra is 0.46\,mm (PBA), 0.40\,mm (measured).

In addition to accurately predicting the penumbra, PBA gives a reasonable account of collimator scattered dose. The long tails of the dose distribution in Fig.\,\ref{fig:HBB0280} come from protons that interact with the opposite face of the collimator, emerge, and cross the beam axis during the 200\,mm drift. These, too, are asymmetric. Those originating in the on-axis face are more plentiful, and PBA reproduces this feature as well.

We can do a rough check of the PBA absolute dose prediction. The fundamental relation
\[\mathrm{dose} = \mathrm{fluence}\times\mathrm{mass\ stopping\ power}\]
leads to
\begin{equation}
D\;=\;\epsilon\;\frac{q}{A}\;\frac{S}{\rho}
\end{equation}
where $D$ (Gy) is dose, $\epsilon$ is the efficiency of the double scattering system (fraction of incident protons reaching the useful field), $q$ (nC) is the incident proton charge, $A$ (cm$^2$) is the area of the useful field and $S/\rho$ (MeV/(g/cm$^2$)) is the mass stopping power. Using the mass stopping power of air at 186\,MeV (the energy of pristine protons reaching the MP) and estimating the efficiency and useful field radius we find
\[D\;=\;0.45\times\frac{1}{\pi\,12^{\,2}}\times4.167\;=\;4.14\quad\mathrm{mGy}\]
per incident nC where PBA finds 4.82\,mGy. Much of that excess is likely genuine, because the seemingly flat central dose includes collimator scattered protons with their considerably higher $S/\rho$.

In proton radiotherapy practice, the dose to the patient ultimately relies on an ADCL calibrated dosimeter placed in a standard field. The dose predicted by PBA should be a good backup check.

\subsection{Discussion}
Unfortunately, there is a problem with the current example. To obtain the excellent results shown, the 60\,mm thick half-beam block was divided into just three 20\,mm slabs. Increasing that to (say) five or ten increases execution time, of course, but also somewhat {\em degrades} agreement with measurement for both the penumbra values and the collimator scattered tails. A well behaved algorithm should improve or level off with increased segmentation: here, three seems to be a magic number. We suspect some sort of interaction between collimator segmentation and the splitting trigger, most likely the $\sigma_x>1$ requirement. 
Whatever the explanation, this property of the algorithm is unacceptable and needs work. 

\section{Example: Comparisons with the Hong Algorithm}

\subsection{Depth Dose}\label{sec:depthDose}
Consider Fig.\,11 (our Fig.\,\ref{fig:urHong11}) of Hong et al. \cite{hong}, the algorithm currently used at the MGH proton radiotherapy center. This test used the 158.6\,MeV contoured scatterer beam at HCL with no range modulation. The terrain
\begin{figure}[h]\begin{verbatim}
terrain: mtl or file, #slabs, thickness (mm) - - - - - - - - - - - - - - -
LEAD            1     0.42
AIR             1    32.
SMALL.MOD       1     1     1 step only, 0.495mm Pb
AIR             5  1162.
REDEF1
02NOV89.CON    32    38.1   HCL contoured scatterer (Fig.5)
AIR             5  4317.
REDEF2
Hong.CIR       40    38.1   brass/air, 2.4mm radius hole
HISTOGRAMS
AIR             1    10.
WATER         100   180.
END           999   999     ==============================================
\end{verbatim}\end{figure}

\noindent is very inefficient for PBA because the open beam has a useful diameter of 240\,mm but only the beam through a 5\,mm diameter hole (ratio of areas $0.04\%$) is used. The PBA calculation for the 1\,cm air gap with two redefinitions, at the collimator surrounding the contoured scatterer and at the collimator under study, takes 2.4 hours. PBA finds only a small fraction of the observed collimator scattered dose (Fig.\,\ref{fig:Hong11}). However, PBA outperforms Hong on the measured difference between the two air gaps. 

Fig.\,\ref{fig:Hong11} shows non-physical fluctuations in collimator scatter, probably linked to non-physical spikes in the $pv$ spectrum (not shown) caused by our convergence criterion Eq.\,\ref{eqn:splitTrigger}.

In a subsidiary study (not shown) we studied the same collimator exposed to a single scattered beam. That requires only one redefinition and is far faster. We were able to match the observed depth-dose quite well by tuning the ellipse parameters of the incident beam, to which collimator scatter appears to be quite sensitive.

\subsection{Transverse Dose}\label{sec:Hong14}
In another test Hong et al. put a 5\,cm Lucite half-beam block in front of a water tank, with a 5.3\,cm air gap between the Lucite and the tank, and performed transverse scans at four depths, with no range modulation. Our Fig.\,\ref{fig:urHong14} reproduces their Fig.\,14. Our Fig.\,\ref{fig:Hong14} is from a PBA run that took 11 hours, a long time. However, this was a drastic test. 

The double scattering system was modeled in full, with redefinitions at the contoured scatterer collimator and at the patient collimator. The `water tank' was divided into one hundred 1.5\,mm slabs with the dose at each evaluated on two axes, 20.4\,K points in all. Starting with one ideal PB, 155\,M PBs were generated of which 35\,M entered the water forming the dose distributions shown. Alternatively, of 966\,pC total (ideally 1000), 26\,pC entered the water, 13\,pC left, and 953\,pC ranged out in a collimator or in water.

Computed entirely from first principles (no source distance or source size, no assumption of constant fluence), our results are somewhat better thsn those of \cite{hong}. Excesses on the LHS in panels (a) and (b), attributed in \cite{hong} to hard scatters (nuclear interactions), are smaller. They are in fact probably related to stopping power. Panel (c), very sensitive to depth, is reproduced by PBA at a single depth, whereas \cite{hong} requires three adjacent depths to account for all the features. Thus, PBA handles range mixing better.

Panel (d) shows a bump in the dose which \cite{hong} attributes to `slit scattering from the thick edge of the Lucite' and which, in consequence, we expected to see in PBA. When we did not, we performed a subsidiary study using a much simpler terrain with pencil-beam scanning to bring out the essence of the interaction with Lucite. (This run, with 80.2\,K PBs evaluated at 20.4\,K points, took 2.2\,min.) A contour plot (Fig.\,\ref{fig:ContLucite}) shows edge scattering very clearly, but the transverse distribution in water on the unblocked side, 1\,cm downstream of where the blocked protons stop, shows no perturbation whatsoever. That agrees with both algorithms, and we now believe that the dose glitch in frame (d) is an experimental artifact.\footnote{~Data were taken using a computer controlled 3D scanning dosimeter. An inadvertent change in depth during the scan would account for the shape of the feature in question.}

\section{Summary}
We have described a deterministic algorithm implemented by a proof-of-principle Fortran program. It is a calculation from first principles, requiring only physical descriptions of the terrain (including the beam line) and of the incident proton beams. It ignores hard scatters, as do other PB algorithms. To that approximation, computed fluence and dose per incident nC are absolute.

\subsection{Parity of Objects}
The terrain is discretized into homogeneous or heterogeneous slabs normal to the nominal beam direction. Heterogeneities are handled by PB redefinition and/or recursive dynamic splitting. All objects in the terrain are discretized and handled the same way. There are no special beam limiting structures. Collimator scatter emerges as a natural consequence, and is a good test of the algorithm.

\subsection{Ur-beams and Pencil Beams}
Incident `ur-beams' and virtual PBs generated during execution have the same mathe\-matical form and are processed the same way. They are described by eleven parameters: a `generation' index $i_g$ and a continuous quantity $q$ for total charge $=q\times0.5^{i_g}$\,nC, five quantities $x$, $x'$, $y$, $y'$ and $j_z$ for position and direction, $pv$ for energy, and three Fermi-Eyges moments (ellipse parameters) for transverse size, divergence and emittance. Cylindrically asymmetric ur-beams would require three additional ellipse parameters. Two ancillary parameters, a serial number and the index of the $z$-plane where the PB was created, are useful in program development.   

\subsection{Parity of Materials}
Each material in the terrain is handled on its own merits, and not referred to water. Thus concepts such as water equivalent depth, radiological path length and residual range in water, are not used. Parameters of the {\O}ver{\aa}s approximation $R=a(pv)^b$ are found from standard range-energy tables for each material encountered, and thereafter characterize that material for energy loss purposes. A scattering power $T_\mathrm{dM}$, also a function of $pv$ and atomic composition, represents MCS. 

\subsection{Stack Processor}
The heart of the algorithm is a computer stack onto which PBs are placed. At each step the current PB is examined relative to the upcoming terrain slice and, depending on the result, either redefined, split, or transported to the next slice. If during transport the PB ranges out, it is deleted from the stack. Otherwise, the contribution of that PB is added to the points of interest in a dose matrix. If the next plane marks the end of the terrain, the PB is then deleted. Otherwise, the loop repeats. When the stack is empty, the next ur-beam is placed on it. After the last ur-beam for that state of the terrain is processed, the next state of the terrain, along with its ur-beams, is loaded. This computational structure is equally suited to scattered beams or to pencil beam scanning.

\subsection{PB Redefinition}
PB redefinition is more or less conventional and, in the present implementation, is triggered (if desired) only at a collimator. The mother PB is replaced by a swarm of daughters, uniformly spaced in an hexagonal array, radiating from the mother's virtual point source, having the same $\sigma_\theta$ as the mother's angular confusion $\theta_\mathrm{c}$, and individual charges that reproduce the mother's Gaussian spatial distribution. PBs that have no chance of emerging from the collimator may be dropped.\footnote{~This is the only vestige of a special `beam defining structure'.} Redefinition parameters are daughter $\sigma_x$, spacing, margin around collimator bore, and the fraction of the mother PB to be redefined if not otherwise limited. 

\subsection{PB Recursive Dynamic Splitting}
PB splitting replaces the mother by seven daughters. One is centered, continues along the mother's trajectory, and has 1/4 of the mother's charge. Six are arranged in a hexagonal array and radiate from the mother's virtual point source, each with 1/8 of the mother's charge. That scheme is inherently 2D and respects, apparently to a sufficiently good approximation, the cylindrical symmetry of MCS in ordinary (non-oriented) materials. As well as charge, the mother's emittance is conserved. There are only two parameters: the ratio $(\approx0.55)$ of  $\sigma_x$ (any daughter) to $\sigma_x$ (mother), and the ratio $(\approx1.15)$ (spatial spread, daughters) to $\sigma_x$ (mother).

The splitting trigger is governed by three parameters: the proximity, in the scale of $\sigma_x$ (mother) of the mother's central axis to a heterogeneity, the absolute value of $\sigma_x$ (mother), and the depth of recursion, that is, $i_g$ (mother). The {\em severity} of the heterogeneity is not considered. That admits virtual heterogeneities (such as an air-to-air boundary), which are useful in program development.

The splitting trigger is, at present, the least satisfactory aspect of the algorithm. In particular, the absolute size requirement seems to have a discontinuous effect on convergence, and interacts with the degree of segmentation into slabs. We are experimenting with dithering the $\sigma_x$ cut as a possible solution. Then the algorithm is no longer strictly deterministic, but the indeterminacy is very small compared to a Monte Carlo. 

\subsection{PB Transport}
Transport of PB parameters from one $z$-plane to the next is conventional. Charge is conserved, $x$ and $y$ trajectories are straight lines, $x'$ and $y'$ are unchanged, $j_z$ is increased by 1, $pv$ is diminished according to the {\O}ver{\aa}s parameters of the current material, and the Fermi-Eyges ellipse parameters $A_0,A_1,A_2$ obey the usual transport equations with $T_\mathrm{dM}(pv)$ as the scattering power. Note that there is no transverse discretization or `voxelizing' of the problem. PB quantities such as $x$ and $y$ can assume any value. Only $z$ is discrete.

\subsection{Dose Accumulation}
Dose accumulation occurs after transport if the new $z$-plane coincides with a measuring plane (MP), that is, a $z$-plane on which we wish to know the fluence or dose. In that case all points of interest in the MP, starting with the one nearest the PB axis and working outwards, are incremented with either the fluence $\Phi$ or the dose $(\Phi\times S/\rho)$ at the appropriate distance from the PB under consideration. $S/\rho$ is an {\em effective} mass stopping power representing the energy loss per g/cm$^2$ of a large cohort of protons, not a single proton. Graphically, it resembles a Bragg peak. PBA offers two options: 

$S_\mathrm{em}/\rho$, a purely EM mass stopping power, is calculated from first principles in the `dose-to' material (air, water, Si $\ldots$) by numerical convolution of its stopping power table with a Gaussian representing range straggling and beam energy spread. $S_\mathrm{em}/\rho$ is appropriate if conditions are such that secondaries from hard scatters can be neglected e.g. collimator scatter in air.

$S_\mathrm{mixed}/\rho$, a mixed mass stopping power, is appropriate in the middle of a large field where hard scatters as well as MCS obey transverse equilibrium. If $S_\mathrm{mixed}/\rho$ is specified, it is obtained from a measurement, either using a small dosimeter on the axis of a large field (conventional Bragg peak measurement) or using a large `pancake' IC to measure a single PB (a so-called integral depth dose or IDD measurement). Usually, the `dose-to' material will be water in this case.

\section{Discussion}
We began this work some years ago to address aesthetic shortcomings of the Hong algorithm.\footnote{~We take this as representative of PB algorithms in current use, and apologize in advance if others are closer to a first-principles calculation.} Protons, of all radiotherapy particles, are the most amenable to dose calculation from first principles. Hard scatters are relatively rare. Good theories of multiple EM interactions (stopping and scattering) are available, and respect the small-angle approximation. Range straggling is modest, so the relation between depth and energy holds up well, as does Fermi-Eyges theory. At our energies, protons are elementary particles with no internal states or fragmentation. Proton radiotherapy is gaining in popularity, providing some motivation.

The Hong algorithm, good performance notwithstanding, is not a solution from first principles. It demands an effective source distance and size, even though these must follow from the scattering system. In other words, it does not solve the whole problem. The fluence in the area of interest is {\em assumed} uniform though that, too, must follow from the scattering system. Collimators are modeled unrealistically. Patient heterogeneities are smeared out in $z$. Their depth acts only via the fraction of rays intercepted. Their stopping and MCS are derived from Hounsfield numbers; thus, they are basically assumed to be water-like.

The sixty-four dollar question\footnote{~In the CBS radio quiz show `Take It or Leave It' (1940-1947) the questions increased in difficulty, you had to decide before each one whether to continue or pocket your winnings, and the last question netted \$64. This led to the popular expression `The \$64 question is $\ldots$'.} is this: given the widespread use of the Hong algorithm, its generally acceptable performance, and the thousands of person-hours expended implementing and testing it, is PBA worth trying to put into practice? It is hard to say, and it will take considerable work to find out. We offer a few comments, in no particular order.

The Hong algorithm is POI-centric. The dose at a particular point can be found very rapidly. That depends on choosing, in advance, PBs headed towards the general vicinity of the POI. However, as we have seen (collimator scatter being an extreme case) it is possible for protons to reach a POI from very different directions with very different stopping powers. Therefore it is hard to see how a more rigorous algorithm can be POI-centric. However, PBA has a countervailing advantage. By lowering the recursion limit, one can get a low-resolution dose image of the entire field very quickly. That may turn out to be useful.

Dose accumulation is an expensive step, and execution time increases with the number of POIs. For instance, in Sec.\,\ref{sec:Hong14} only four MPs (808 POIs) were actually necessary, and execution time could have been far shorter. In PBA the association of MPs with $z$-planes (slabs) is fundamental since PBs are only `known' on $z$-planes. However the requirement that MPs correspond to {\em contiguous} $z$-planes could easily be lifted, decoupling them from $z$-segmentation (that is, transport accuracy).

It goes without saying that PBA already makes extensive use of lookup tables, packaged as functions with initialization and execution entries. For instance,
\begin{quote}\begin{verbatim}i = InitSpwrTabl(t1,pv1)\end{verbatim}\end{quote}
sets up, for each listed material, a table of $T_\mathrm{dM}$ (mrad$^2$/mm) for incident energy $T_1$ corresponding to $(pv)_1$. Subsequently
\begin{quote}\begin{verbatim}spwr = SpwrTabl(r,k)\end{verbatim}\end{quote}
returns $T_\mathrm{dM}$ corresponding to residual range $r$ in material number $k$. We often use linear interpolation with constant spacing. That may be inefficient memory-wise, but is very fast because finding the table position only requires one floating-point multiply.

The major task still required (after rewriting PBA in a more fashionable language) is bringing in patient information from CT. The Hounsfield map must be re-sliced perpendicular to the nominal beam direction for each field. Structures must be contoured, and their atomic composition defined by feeding in anatomical information. Presumably that is all within the reach of modern image processing techniques.

Once the atomic composition, however complicated, is known, the formula 
\begin{equation}\label{eqn:Rasouli}
\frac{1}{\rho R}\;=\;\sum_i\frac{w_i}{\rho_i R_i}
\end{equation}
yields a very good approximation to the mass range $\rho R$ (g/cm$^2$) at any energy \cite{Rasouli2015}. $w_i$ is the fraction by weight of the $i^\mathrm{th}$ atomic constituent and $\rho_iR_i$ its mass range at the same energy. This avoids having to compute and integrate the stopping power of the compound or mixture, and Janni\,82 \cite{janni82} gives the range-energy relation for all atomic elements. One need only apply Eq.\,\ref{eqn:Rasouli} at two fiducial energies to obtain the two {\O}ver{\aa}s parameters. (Conceivably a more direct connection between Hounsfield numbers and {\O}ver{\aa}s parameters will be found.)

As Sec.\,\ref{sec:HCLdata} makes clear, speed depends critically on the sharpest dose feature that needs to be reproduced. If, instead of requiring the dose 0.6\,mm from the edge, we are content with 10\,mm, PBA is enormously faster. Immediately behind (say) a titanium surgical implant in the patient, the dose may have a large gradient, but this is very rapidly washed out by angular confusion, and may not be clinically important.

In closing, we note that pencil-beam scanning is of greater interest nowadays than the passively scattered examples we used (in order to exploit existing data). Our algorithm solves, in either mode, heretofore neglected problems such as collimator scatter and surgical implants. In addition, in scanned beams it  may well be faster, more accurate, and easier to commission than algorithms in current use. It would be interesting to find out.

\section{Acknowledgements}
We are indebted to Harvard University, the Physics Department, and the Laboratory for Particle Physics and Cosmology 
for continuing support. The papers of Nobuyuki Kanematsu provided the seminal idea and many other insights. We thank David Hall for the TOPAS simulation. We thank Hanne Kooy for sharing the Burr Center commissioning data, and Damien Prieels, Victor Breev, Bob Brett, Ethan Cascio, Martijn Engelsman and Hsiao-Ming Lu for helping us determine dimensions of the MGH beam spreading system as built. 

\appendix
\section{TOPAS Monte-Carlo Details}\label{sec:TOPAS}
This is David Hall's description, slightly edited, of the TOPAS Monte Carlo run in connection with our first example, Sec.\,\ref{sec:HCLdata}.
\begin{quote}
`The MC simulation was performed using TOPAS version 1.3 \cite{Perl2012}, which is based on Geant4 version 9.6.p04 \cite{Agostinelli2003}. $2\times10^9$ proton histories were simulated, originating 1mm upstream of the lead scatterer. TOPAS default physics settings, as validated by Testa et al. \cite{Testa2013} were used, which correspond to a physics list containing the Geant4 modules: g4em-standard\_opt3, g4h-phy\_QGSP\_BIC\_HP, g4decay, g4ion-binarycascade, g4h-elastic\_HP, g4q-stopping and g4radioactivedecay. The range threshold for secondary particle production was 0.05mm. At each measurement plane in the air following the collimator, dose-to-silicon was scored in a set of concentric tubes centered on the beam axis. Each tube had length 0.2mm and radial width 0.4mm. The scored dose in each tube was ascribed to a radial distance according to the following formula, which accounts for the bias towards larger radii due to the expanding volumes involved:
\[r_\mathrm{MC}\;=\;(2/3) * (r_\mathrm{outer}^3 - r_\mathrm{inner}^3) / (r_\mathrm{outer}^2 - r_\mathrm{inner}^2)\]
where $r_\mathrm{inner}$ and $r_\mathrm{outer}$ are the inner and outer radii of the tube in question.'

`My simulated brass is 70/30 Cu/Zn with density 8.550 g/cm$^3$, with mean excitation energy 324.4 eV.'

`Your questions about which physics processes are included in which Geant4 physics modules are answered in the Physics Reference Manual. The document is huge, but I think that I have an answer now. It is as you guessed, with the Moli\`ere tail being included in my `EM only' results. In fact, in Geant4 versions before 10.2, these interactions are handled by the MCS process. After version 10.2, they are handled by a separate single scattering process. Anything involving the nuclear strong force is going to be handled by nuclear processes and is not included in my `EM only` results.'

`I ran 200 parallel simulations of 10M histories on RHEL6 nodes. Each takes around 20 hours (TOPAS 1.3 runs single-threaded only).'
\end{quote}

\section{Point-Source Approximation and Dose $z$-Dependence}\label{sec:inverseRsq}

We inquire here whether the inconsistency between $D_1(0)$ and $D_6(0)$ could possibly result from assuming a virtual point source, rather than an extended source, when redefining a PB. Put more precisely, what is the ratio between the actual fluence on axis to that from a virtual point source? That can be found in closed form for a beam line consisting (as ours does) of uniform slabs. See \cite{transport2012} for notation and further discussion of the formulas used below. 

Consider an arbitrary stack of homogeneous slabs upstream of a {\em defining plane} MP$_0$, at which we take $z=0$. We need not specify the stack or the Gaussian beam entering it other than to say that the FE moments $A_0(0)$, $A_1(0)$ and $A_2(0)$ are known, as (therefore) are related quantities such as $S_\mathrm{vir}$, $S_\mathrm{eff}$ and $x_\mathrm{eff}$. We locate MP$_0$ at the {\em upstream} collimator face because that is where the incident PB is redefined. 

\begin{figure}[p] 
  \centering
  \includegraphics[bb=20 40 576 469,width=4.54in,height=3.5in,keepaspectratio]{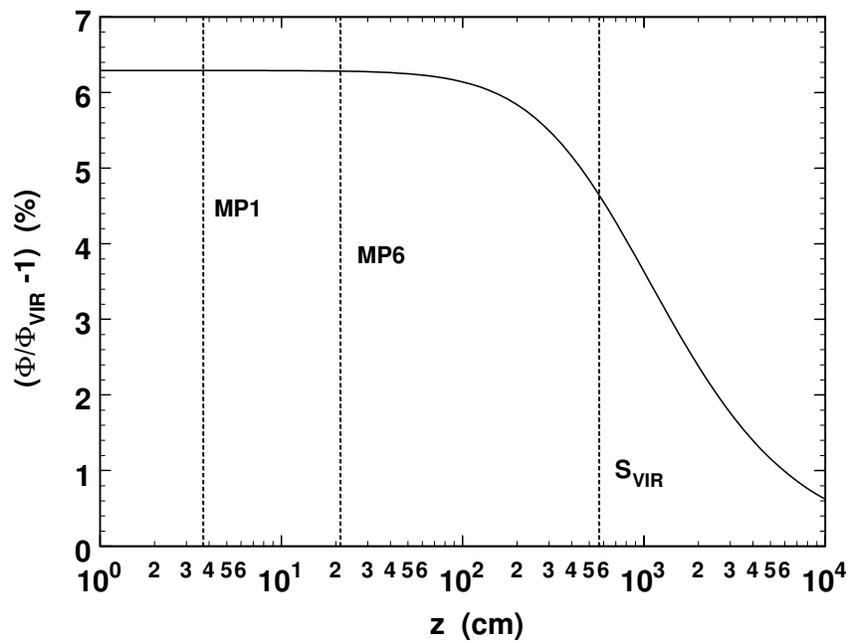}
  \caption{$(\Phi(z)/\Phi_\mathrm{vir}(z)-1)$ (\%) vs. distance $z$ (cm) between the defining plane MP$_0$ and the measuring plane using the parameters of Sec.\,\ref{sec:HCLdata}.}
  \label{fig:fluenceCorr}
\end{figure}

Now let us compute the on-axis fluence $\Phi(z)$ at a second MP at $z\ge0$. The 1D fluence per proton is found by integrating
\begin{equation*}\label{eqn:Pxt}
P(x,\theta)\,dx\,d\theta\;=\;\frac{1}{2\pi\sqrt{B}}\;e^{\displaystyle{
  -\,\frac{1}{2}\frac{A_0x^2-2A_1x\theta+A_2\theta^2}{B}}}\,dx\,d\theta
\end{equation*}
over $\theta$ with the result
\begin{equation*}\label{eqn:Px}
P(x)\;=\;\frac{1}{\sqrt{2\pi A_2}}\;e^{\displaystyle{-\frac{1}{2}\frac{x^2}{A_2}}}
\end{equation*}
Multiplying this by a similar expression for the orthogonal coordinate $y$, assuming the $A_i(z)$ to be equal for $x$ and $y$, and setting $x=y=0$ we find the on-axis 2D fluence (1/cm$^2$) per proton
\begin{equation*}
\Phi(z)\;=\;\frac{1}{2\pi A_2(z)}
\end{equation*}
Using the usual transport relation for $A_2$ and ignoring scattering in air downstream of MP$_0$
\begin{equation*}
\Phi(z)\;=\;\frac{1}{2\pi(A_2(0)+2A_1(0)z+A_0(0)z^2)}
\end{equation*}
The same formula, if $A_1\doteq A_2\doteq 0$, gives the fluence from a virtual point source a distance $S_\mathrm{vir}$ upstream of MP$_0$
\begin{equation*}
\Phi_\mathrm{vir}(z)\;=\;\frac{1}{2\pi A_0(0)(z+S_\mathrm{vir})^2}
\end{equation*}
Taking the ratio and simplifying we have finally
\begin{equation}
\frac{\Phi(z)}{\Phi_\mathrm{vir}(z)}\;=\;\frac{A_0(z+S_\mathrm{vir}^2)}{A_0(z+S_\mathrm{eff}^2)+x_\mathrm{eff}^2}
\end{equation}
where all quantities on the RHS except $z$ itself are evaluated at MP$_0$. The ratio correctly equals 1 for $z$ very large or in the trivial case where the emittance at the defining plane is 0. Its deviation from 1 is plotted in Fig.\,\ref{fig:fluenceCorr} using the parameters of Sec.\,\ref{sec:HCLdata}. The maximum effect is only $6.3\%$ and is the same at MP1 and MP6, so it cannot account for the inconsistency in dose at those locations.

\clearpage

\clearpage
\listoffigures

\clearpage

\begin{figure}[p] 
  \centering
  \includegraphics[bb=0 0 468 360,width=4.55in,height=3.5in,keepaspectratio]{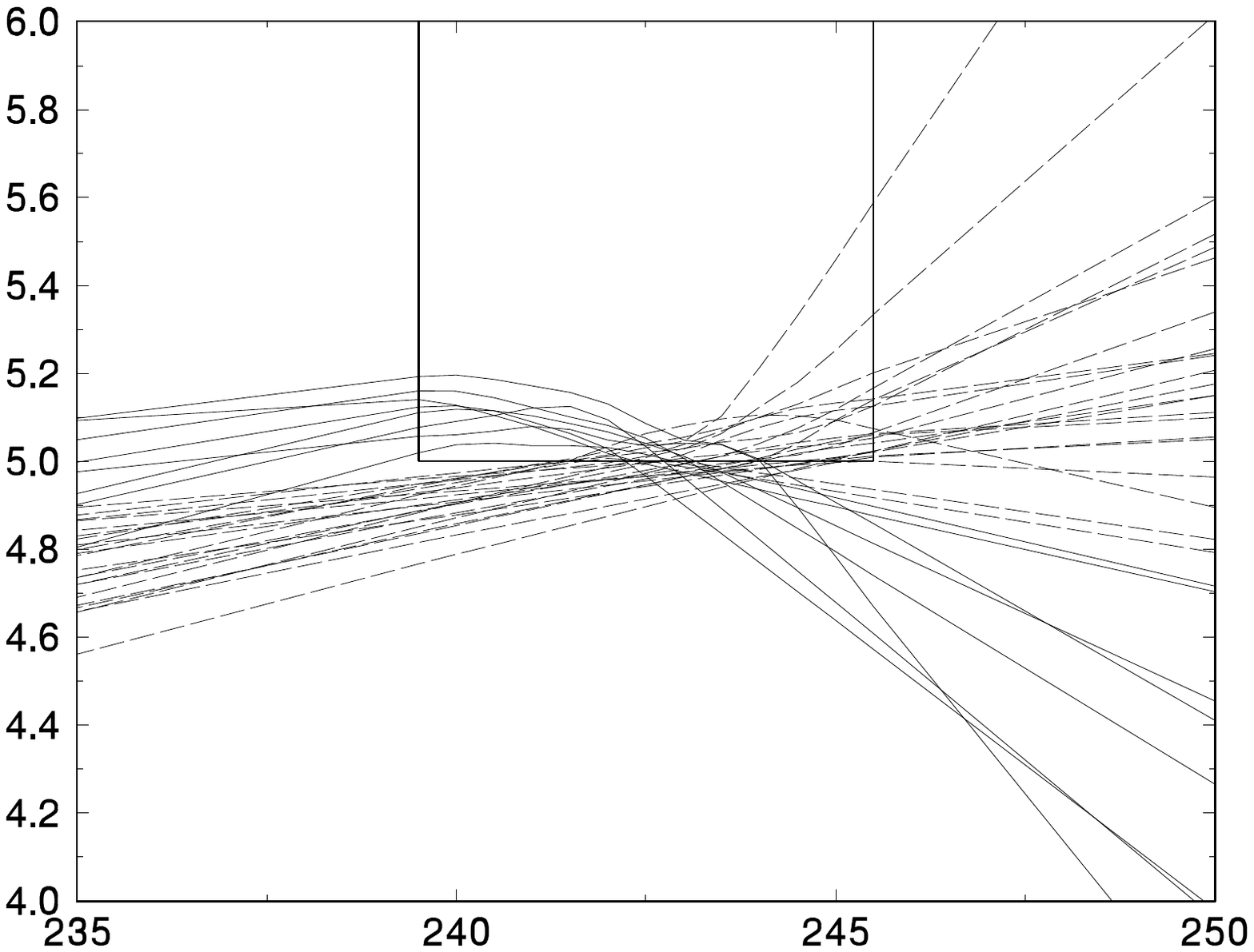}
\caption{Monte Carlo simulation of typical collimator scatter in brass (geometry different from 
Fig.\,\ref{fig:HCLX0258}). Dimensions are in cm, the transverse scale is greatly exaggerated, and only protons that lose energy  in the brass (but emerge) are shown.}
\label{fig:GMCTR2}
\end{figure}

\begin{figure}[p] 
  \centering
  \includegraphics[bb=20 20 592 449,width=4.67in,height=3.5in,keepaspectratio]{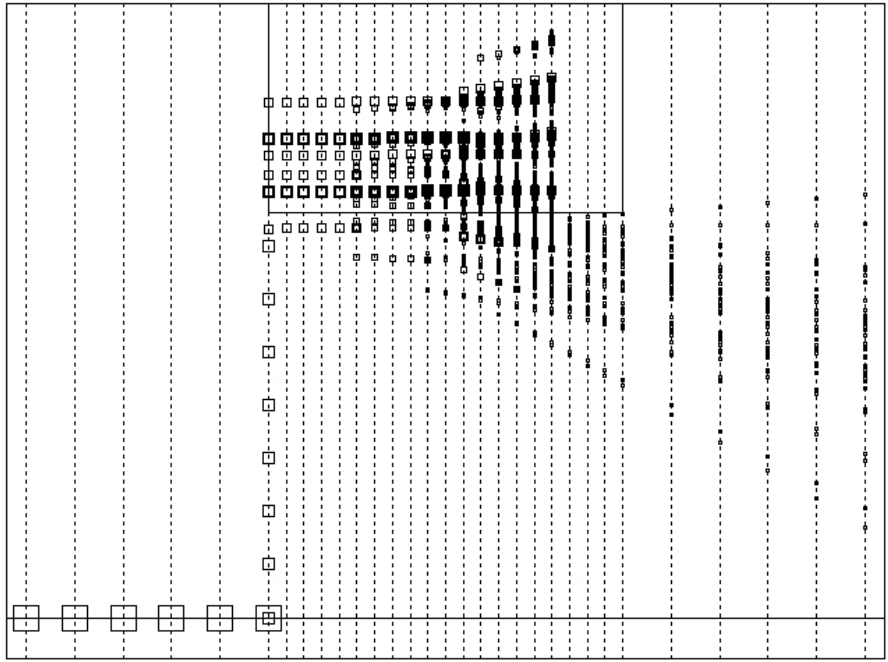}
  \caption{Side view of PB tracks generated by the algorithm. Box sizes vary as $\log_{10}$ of the charge carried by that PB. From left: ur-beam; regular array of PBs from redefinition at collimator face; daughters of dynamically split PBs. For visual effect, many tracks are suppressed.}
  \label{fig:PBtracks}
\end{figure}

\begin{figure}[p] 
  \centering
  \includegraphics[bb=0 0 640 480,width=4.67in,height=3.5in,keepaspectratio]{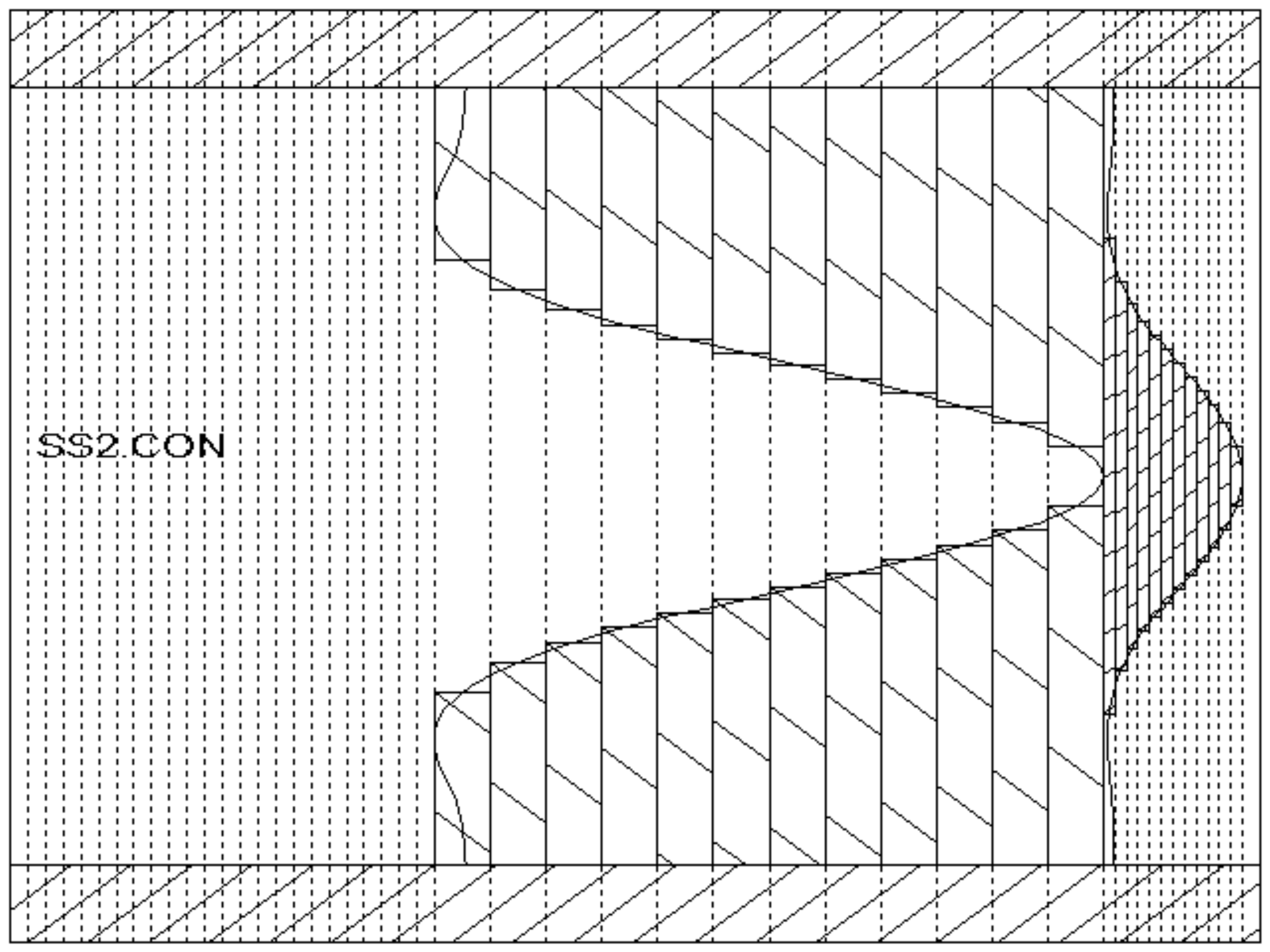}
  \caption{A compensated contoured scatterer (Lucite and lead) with surrounding brass collimator, divided into slabs.}
  \label{fig:SS2}
\end{figure}

\begin{figure}[p] 
  \centering
  \includegraphics[bb=20 40 576 469,width=4.54in,height=3.5in,keepaspectratio]{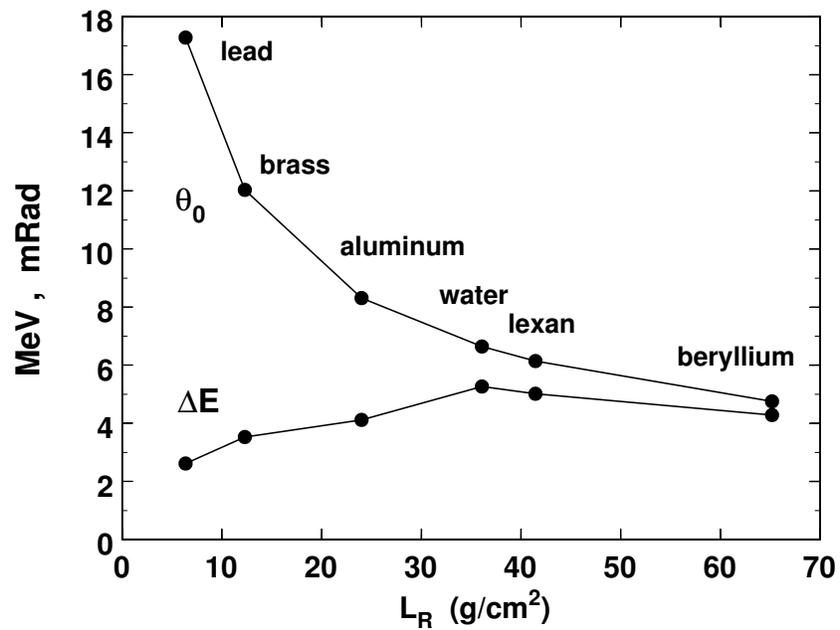}
  \caption{Multiple scattering angle and energy loss for 160\,MeV protons traversing 1\,g/cm$^2$ of various materials \cite{icru49}.}
  \label{fig:hiZloZ}
\end{figure}

\clearpage

\begin{figure}[p] 
  \centering
  \includegraphics[bb=20 20 592 424,width=4.95in,height=3.5in,keepaspectratio]{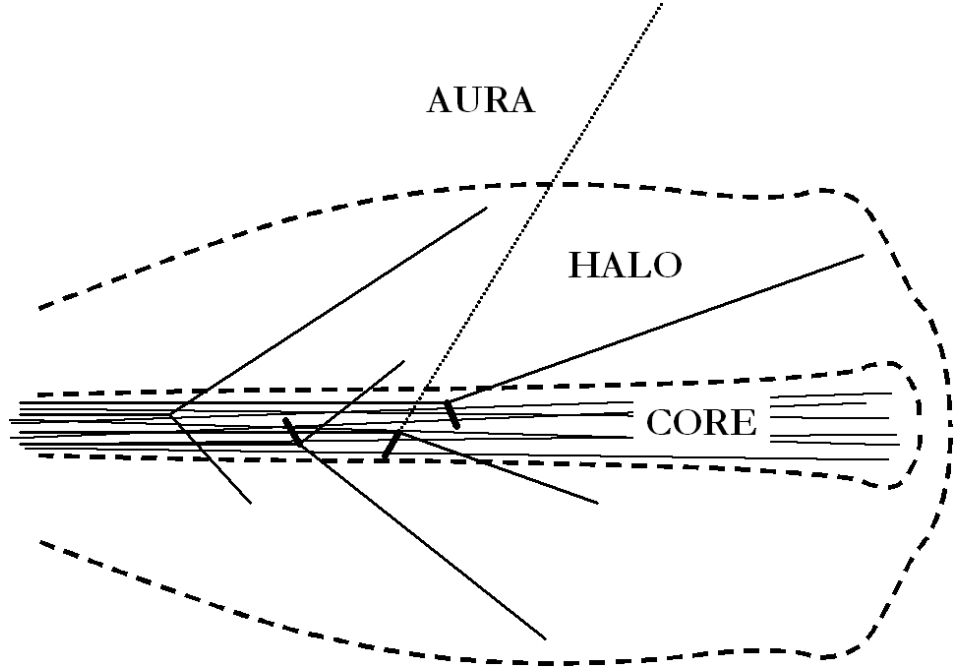}
  \caption{Core, halo and aura with schematic reactions (from left) $^1$H(p,p)p, $^{16}$O(p,2p)$^{15}$N, $^{16}$O(p,pn)$^{15}$O and $^{16}$O(p,p)$^{16}$O. Recoil nuclei ranges are exaggerated. The dashed lines are 10\% and 0.01\% isodoses drawn to scale.}
  \label{fig:haloReactions}
\end{figure}

\begin{figure}[p] 
  \centering
  \includegraphics[bb=20 20 592 603,width=3.44in,height=3.5in,keepaspectratio]{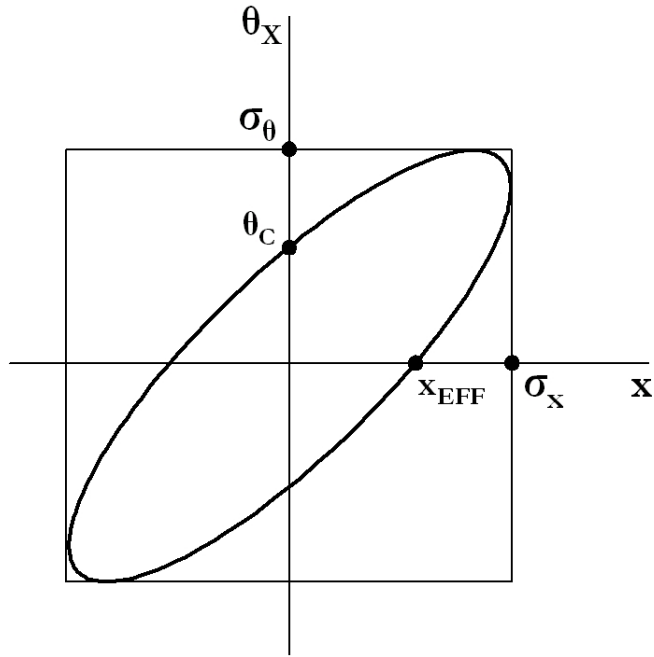}
  \caption{Convenient ellipse parameters are $\sigma_x$, $\sigma_\theta$ and $\theta_C$.}
  \label{fig:ellipseThetaC}
\end{figure}

\begin{figure}[p] 
  \centering
  \includegraphics[bb=0 0 640 480,width=4.67in,height=3.5in,keepaspectratio]{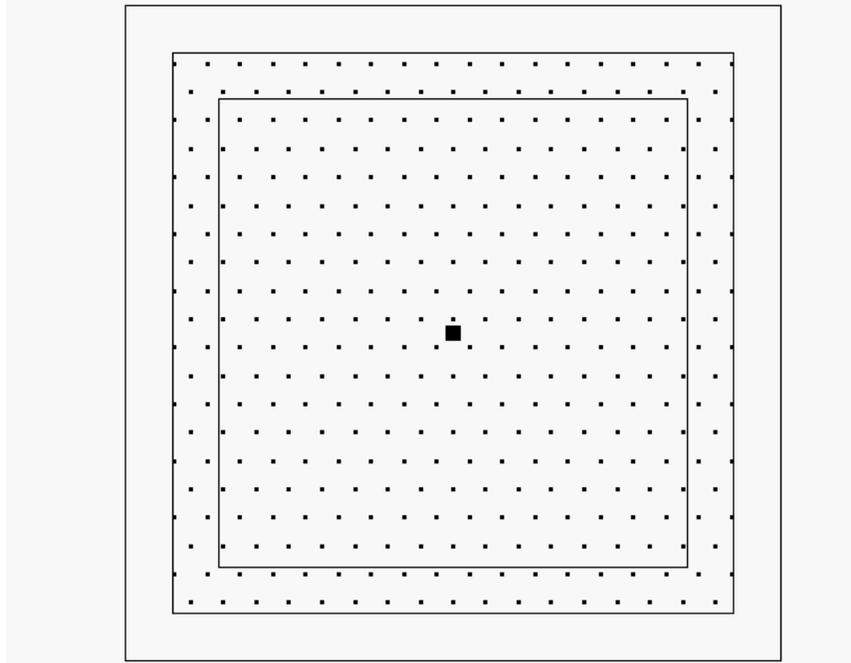}
  \caption{Redefined PB in $xy$ plane showing daughter centroids. The innermost square is the hole in the collimator and a 1\,mm margin is furnished.}
  \label{fig:RedXY}
\end{figure}

\begin{figure}[p] 
  \centering
  \includegraphics[bb=0 0 640 480,width=4.67in,height=3.5in,keepaspectratio]{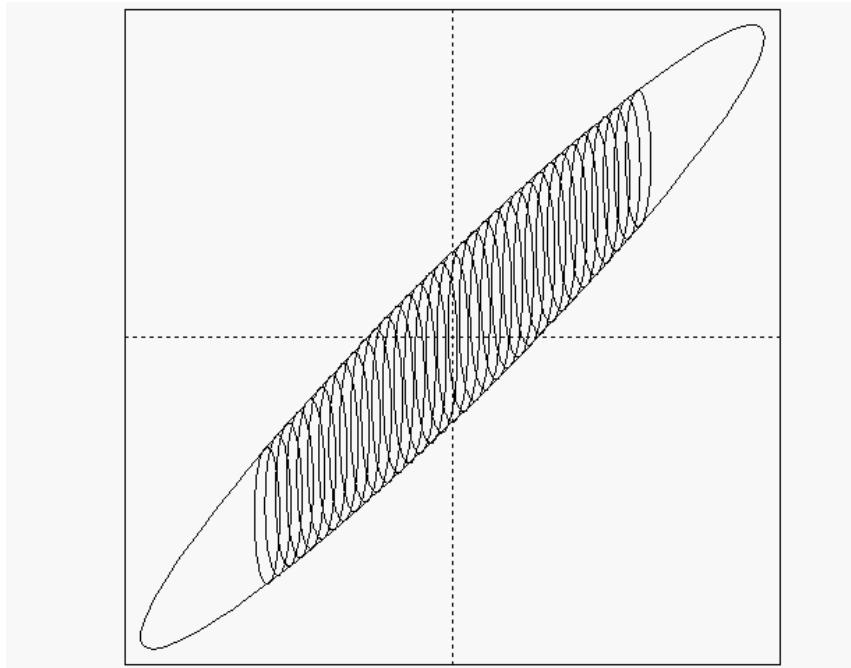}
  \caption{Redefined PB in phase space. Daughter ellipses actually have $\sqrt{A_0}$ equal to $\theta_\mathrm{C}$ of the mother, but are adjusted for the same phase space density at the point of tangency.}
  \label{fig:RedPS}
\end{figure}

\clearpage

\begin{figure}[p] 
  \centering
  \includegraphics[bb=0 0 640 480,width=4.67in,height=3.5in,keepaspectratio]{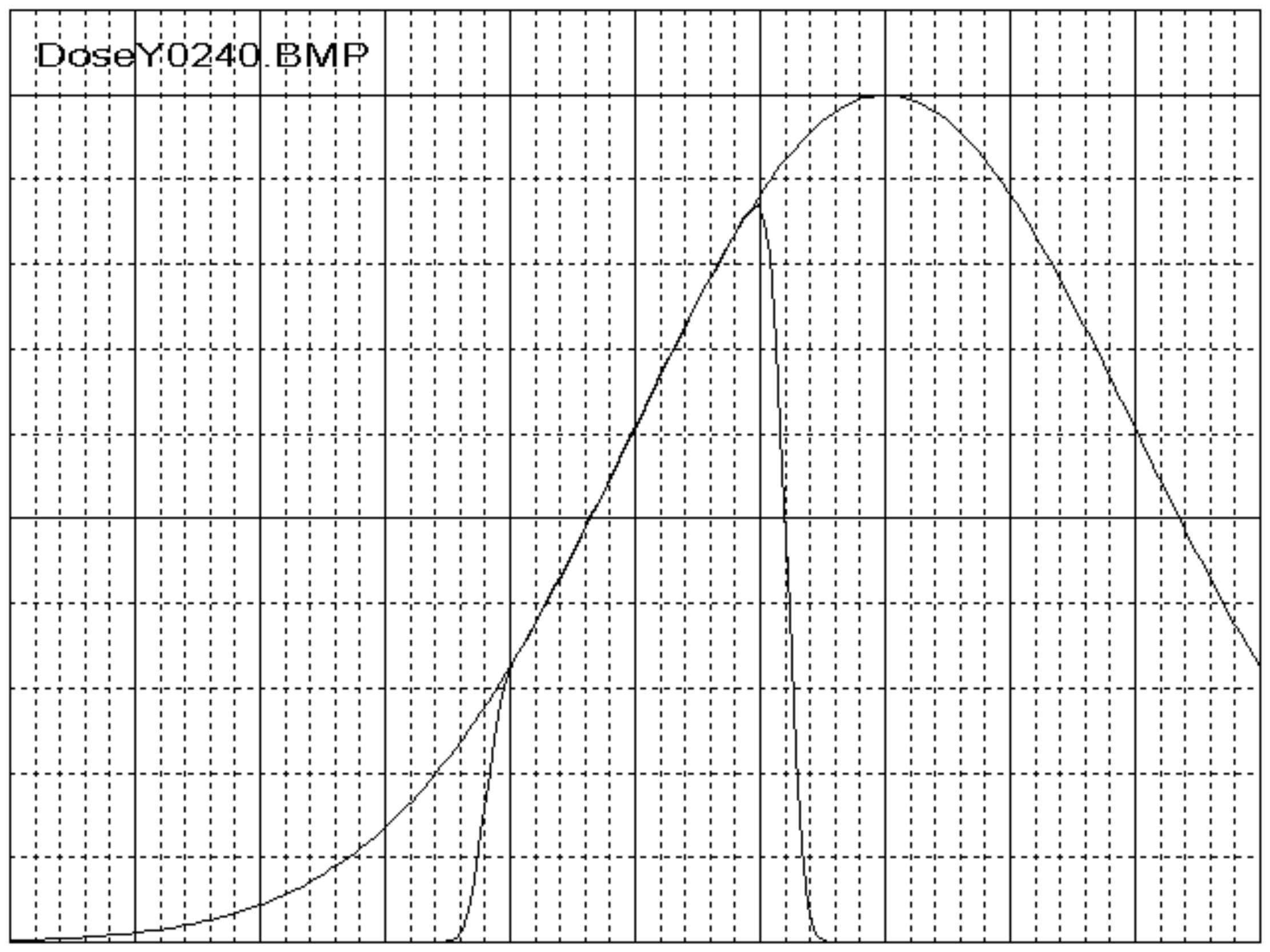}
  \caption{Gaussian compliance to a mother PB, offset in the $y$ direction, at a collimator.}
  \label{fig:RedComp}
\end{figure}

\begin{figure}[p] 
  \centering
  \includegraphics[bb=20 20 592 449,width=4.67in,height=3.5in,keepaspectratio]{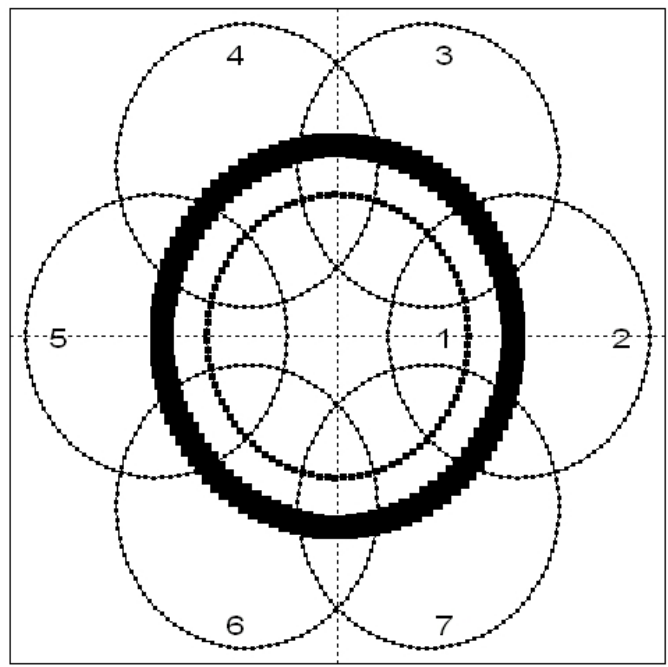}
  \caption{Split mother and daughters shown in $xy$ plane. Line weight is proportional to charge (8, 2, 6$\times1$). Off-axis daughters radiate from mother's virtual point source. Daughter numbers correspond to Sec.\,\ref{sec:seven}}
  \label{fig:SplitXY}
\end{figure}

\begin{figure}[p] 
  \centering
  \includegraphics[bb=20 20 592 449,width=4.67in,height=3.5in,keepaspectratio]{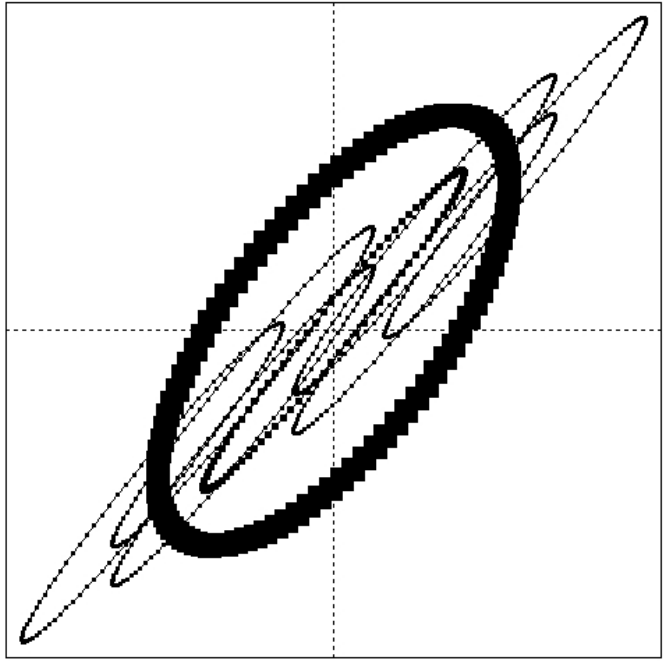}
  \caption{Split mother and daughters shown in phase space. Line weight is proportional to charge (8, 2, 6$\times1$). Daughters radiate from mother's virtual point source. Total daughter emittance = mother emittance. Four daughters are displaced vertically for visual effect.}
  \label{fig:SplitPS}
\end{figure}

\begin{figure}[p] 
  \centering
  \includegraphics[bb=20 20 592 449,width=4.67in,height=3.5in,keepaspectratio]{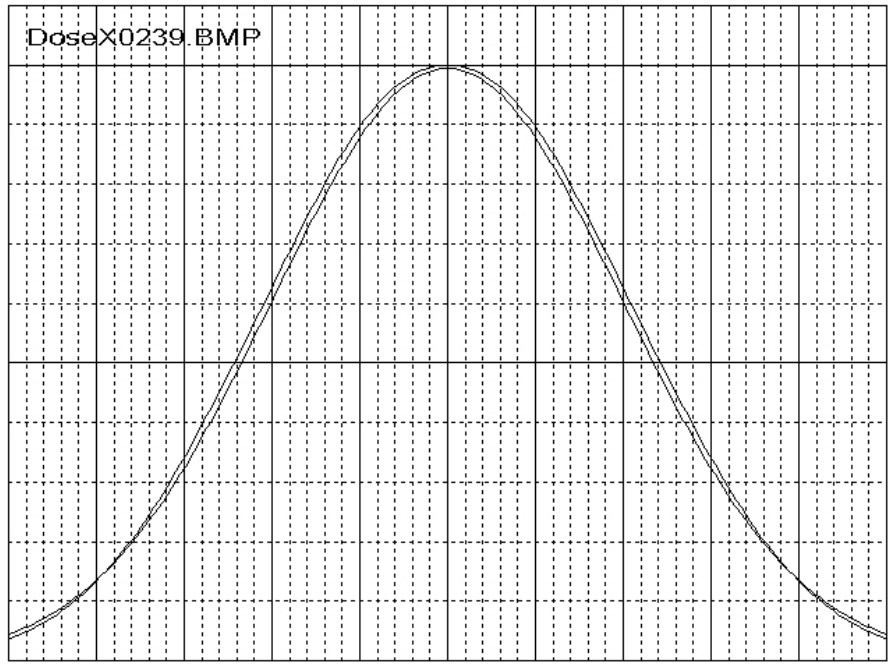}
  \caption{Gaussian compliance immediately after splitting. The dose distribution, along the $x$ axis, of the original beam is compared with the sum of 751 PBs (produced by recursive splitting) transported through 0.1\,mm of air. A comparison along the $y$ axis is similar.}
  \label{fig:SplitX}
\end{figure}

\clearpage

\begin{figure}[p] 
  \centering
  \includegraphics[bb=20 20 592 461,width=4.54in,height=3.5in,keepaspectratio]{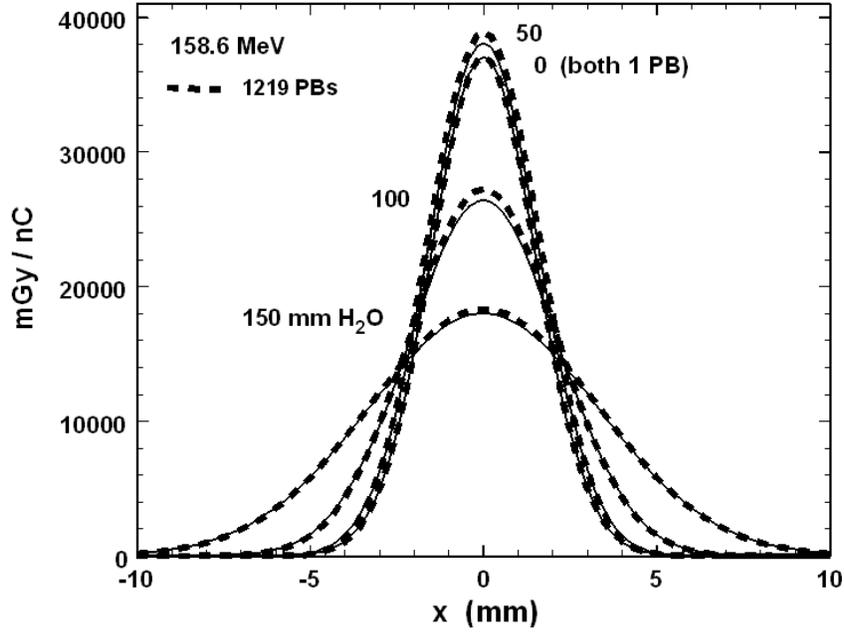}
  \caption{Gaussian compliance after splitting and transport. The dose distribution, along the $x$ axis, of the original beam is compared with the sum of 1219 PBs (produced by recursive splitting) transported through 50, 100 and 150\,mm (near stopping thickness) of water. A comparison along the $y$ axis is similar.}
  \label{fig:HMLUtest}
\end{figure}

\begin{figure}[p] 
  \centering
  \includegraphics[bb=20 20 592 465,width=4.5in,height=3.5in,keepaspectratio]{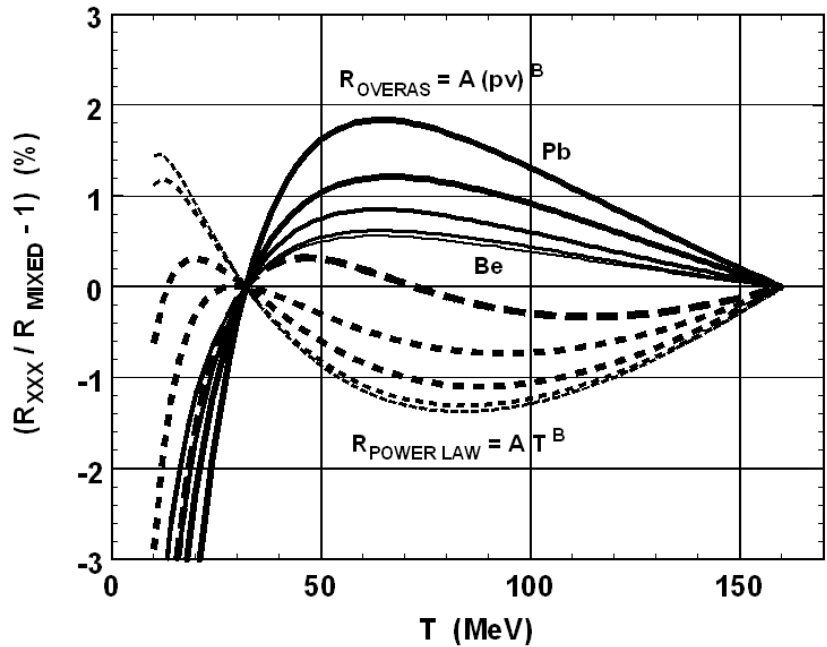}
  \caption{Comparison of the {\O}ver{\aa}s range-energy approximation (solid lines) to the more common $R=aT^b$ (dashed lines). Materials in order of increasing line weight are Be, water, Al, Cu, Pb. Parameters in this example are computed from exact fits at 32 and 160\,MeV. The MIXED range-energy table corresponds to ICRU\,49 \cite{icru49} except for water which uses Janni\,82 \cite{janni82}.}
  \label{fig:TestRover}
\end{figure}

\begin{figure}[p] 
  \centering
  \includegraphics[bb=20 40 576 469,width=4.54in,height=3.5in,keepaspectratio]{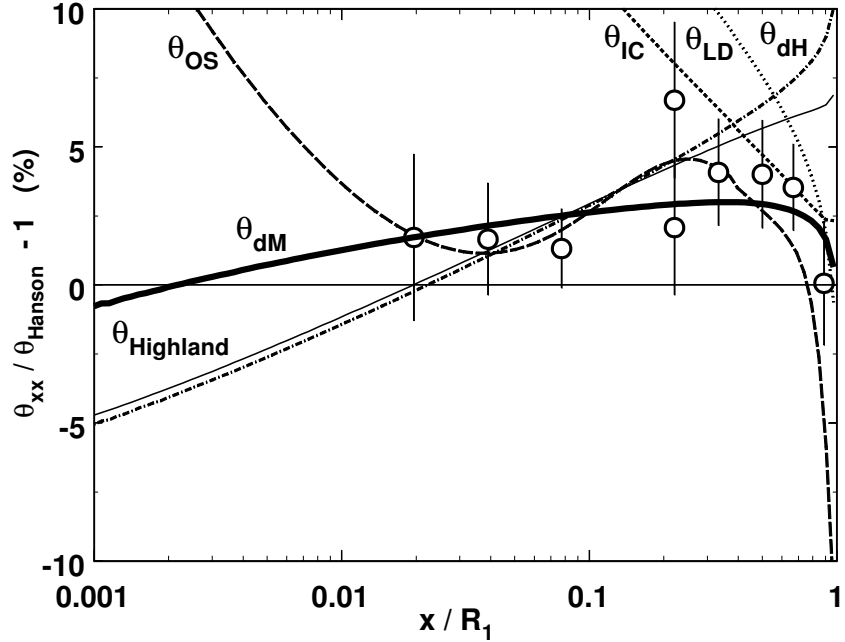}
  \caption{For polystyrene, deviation of $\theta_0$ from $\theta_\mathrm{Hanson}$, according to various formulas, at $T_1=158.6$\,Mev (experimental points from \cite{mcsbg}). Each $\theta$ is the integral of the corresponding $T_\mathrm{xx}$ except $\theta_\mathrm{Highland}$ which is from the generalized Highland formula.}
  \label{fig:ExptFigPoly}
\end{figure}

\begin{figure}[p] 
  \centering
  \includegraphics[bb=20 40 576 469,width=4.54in,height=3.5in,keepaspectratio]{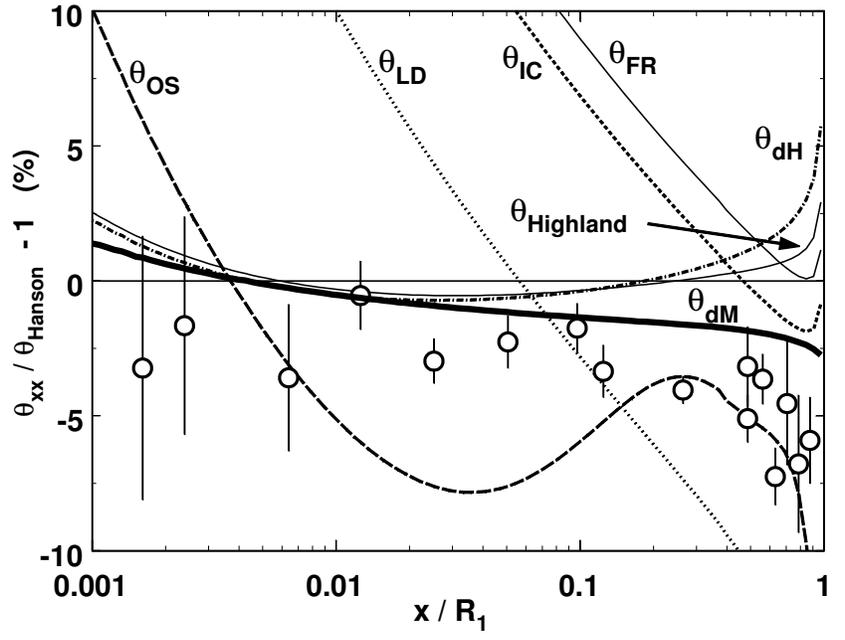}
  \caption{Same as Figure \ref{fig:ExptFigPoly} for lead.}
  \label{fig:ExptFigLead}
\end{figure}

\clearpage

\begin{figure}[t]
\begin{verbatim}
         DO jTerr = np1,np2  !  states of terrain
C
       CALL InitWhatsHere(inpFile,'terrain:',jTerr,wgt,jzhist,jzred)
       CALL InitUrbeams(nrows,pvex,scanHW,zSource,urp,sigt,thtc,sigx)
C
           DO jUrb = 1,nUrb  !  ur-beams
       CALL PushUrbeam(jUrb)
       rangeOut = .FALSE.
       IF (WantDose(rangeOut)) CALL AddToDose(idDoseto,jTerr,bpf)
C
             DO              !  stack
       CALL WhatsHere(xbu(k2),ybu(k2),jzbu(k2),kml,dzz,rho,scl,dsqnh)
C
               IF (red(1)) THEN 
       CALL Redefine(showRed(1),redp(1,1),ml(jzred(1)),stopnC)
               ELSE IF (red(2)) THEN
       CALL Redefine(showRed(2),redp(1,2),ml(jzred(2)),stopnC)
               ELSE IF (WantSplit(dsqnh,jzred)) THEN
       CALL Split(snum)
               ELSE
       CALL Transport(kty(jzbu(k2)),kml,dzz,dsqnh,rho,rangeOut,reachEnd)
       IF (WantDose(rangeOut)) CALL AddToDose(idDoseto,jTerr,bpf)
                 IF (reachEnd.OR.rangeOut) THEN
       k2 = k2 - 1
                 END IF
               END IF  !  redefine, split or transport
C
       IF (k2.LE.0) EXIT
C
             END DO          !  stack
C
           END DO            !  ur-beams
C
         END DO              !  states of terrain
\end{verbatim}
\caption{Basic program.\label{fig:basicProgram}}
\end{figure}

\clearpage
\begin{figure}[p] 
  \centering
  \includegraphics[bb=0 11 576 432,height=4.5in,width=6.in]{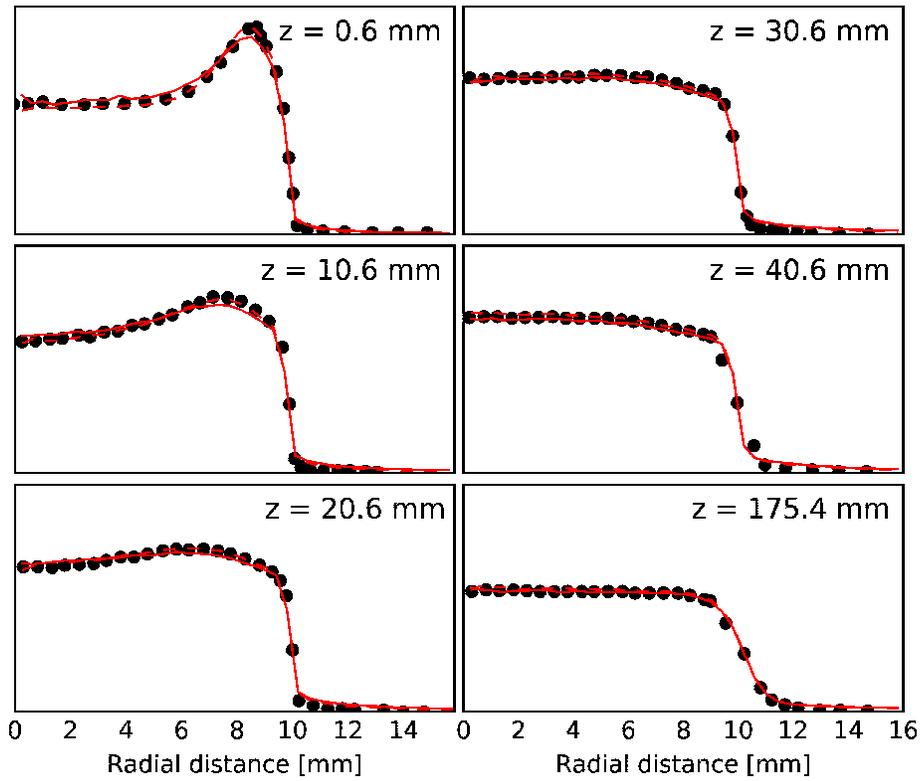}
  \caption{TOPAS simulation (courtesy David Hall) of the HCL data. Points: HCL data; solid line: TOPAS, all interactions; dashed line: TOPAS, EM only (including Moli\`ere tail). A single normalization constant for all TOPAS lines and all six MPs is calculated to fit the first few points $D_6(\approx0)$ in MP6 (175.4\,mm).}
  \label{fig:TOPAS}
\end{figure}

\clearpage
\begin{figure}[p] 
  \centering
  \includegraphics[bb=0 0 640 480,width=4.67in,height=3.5in,keepaspectratio]{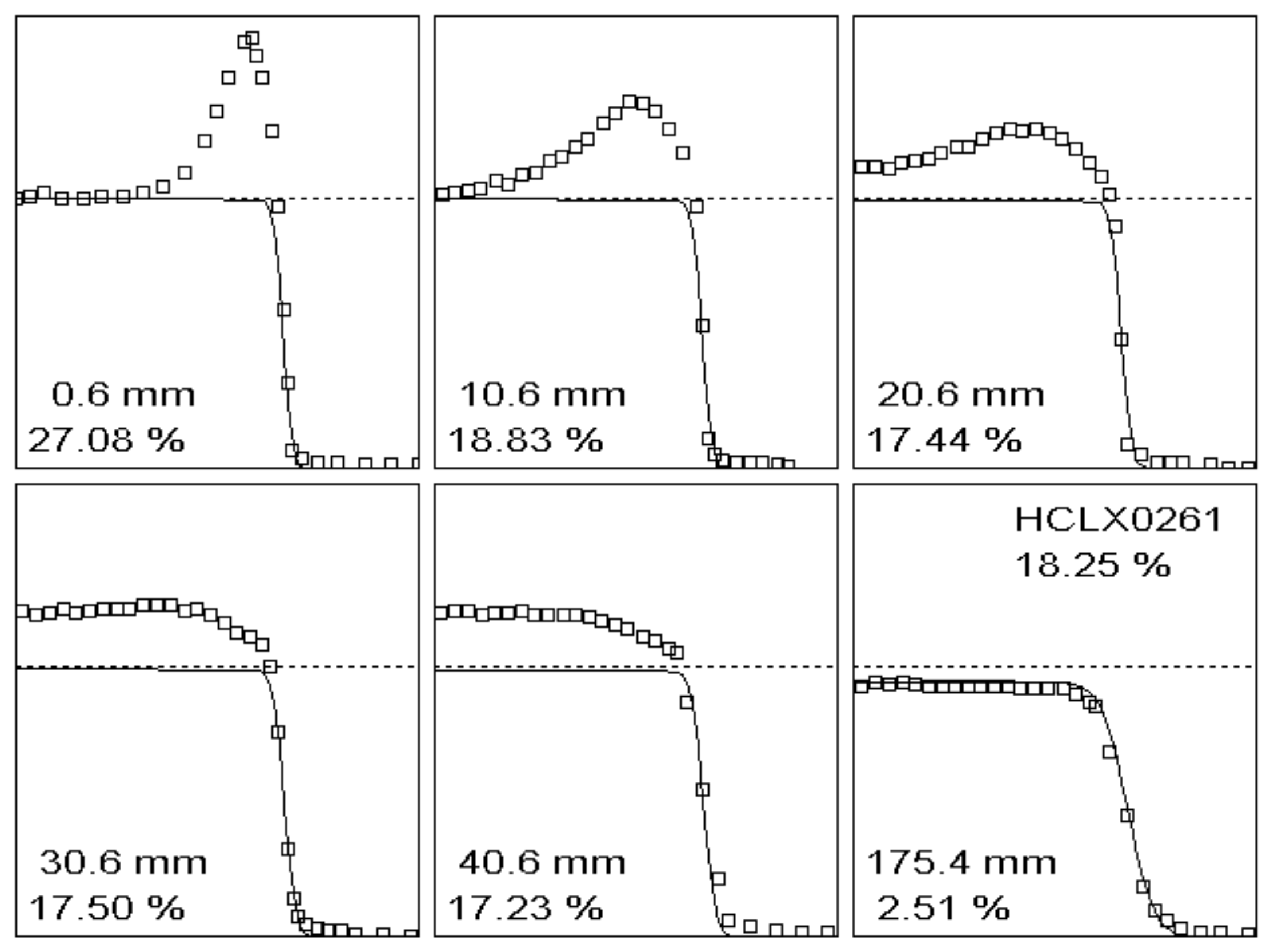}
  \caption{Points: HCL data; line: PBA calculation similar to Fig.\,\ref{fig:HCLX0258} below, but dose accumulated only for pristine protons ($pv>284$\,MeV). A single normalization constant for all PBA results is calculated to fit the first few points $D_1(\approx0)$ in MP1, $z=0.6$\,mm.}
  \label{fig:HCLX0261}
\end{figure}

\begin{figure}[p] 
  \centering
  \includegraphics[bb=20 40 576 469,width=4.54in,height=3.5in,keepaspectratio]{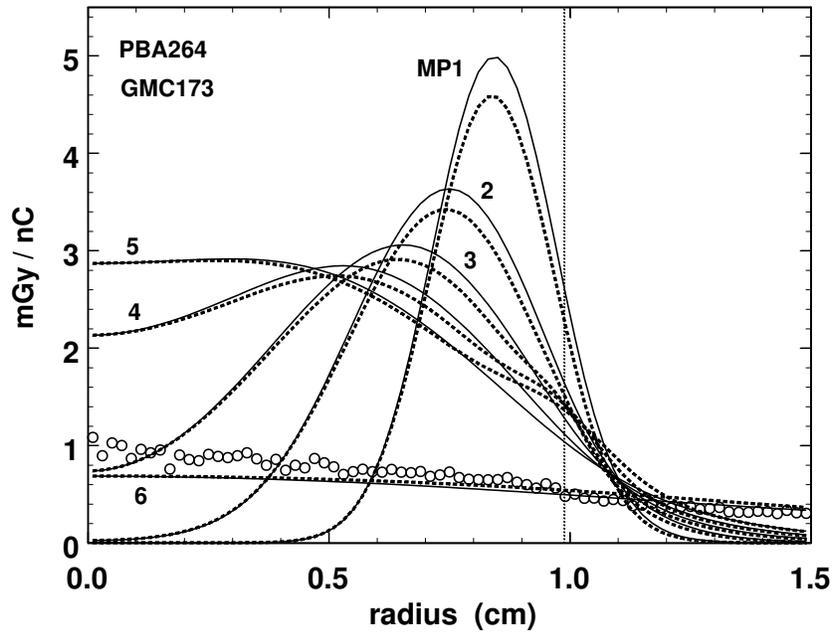}
  \caption{PBA run similar to Fig.\,\ref{fig:HCLX0262} with dose accumulated only for degraded protons, $0<pv<284$\,MeV. Solid lines: dose along $x$ axis; dashed lines, dose along $y$ axis. Vertical line: geometric edge of collimator. Points: simulation with a home-made Monte Carlo program GMC, 100\,M events. All doses are absolute (no normalization).}
  \label{fig:HCL0264}
\end{figure}

\begin{figure}[p] 
  \centering
  \includegraphics[bb=0 0 640 480,width=4.67in,height=3.5in,keepaspectratio]{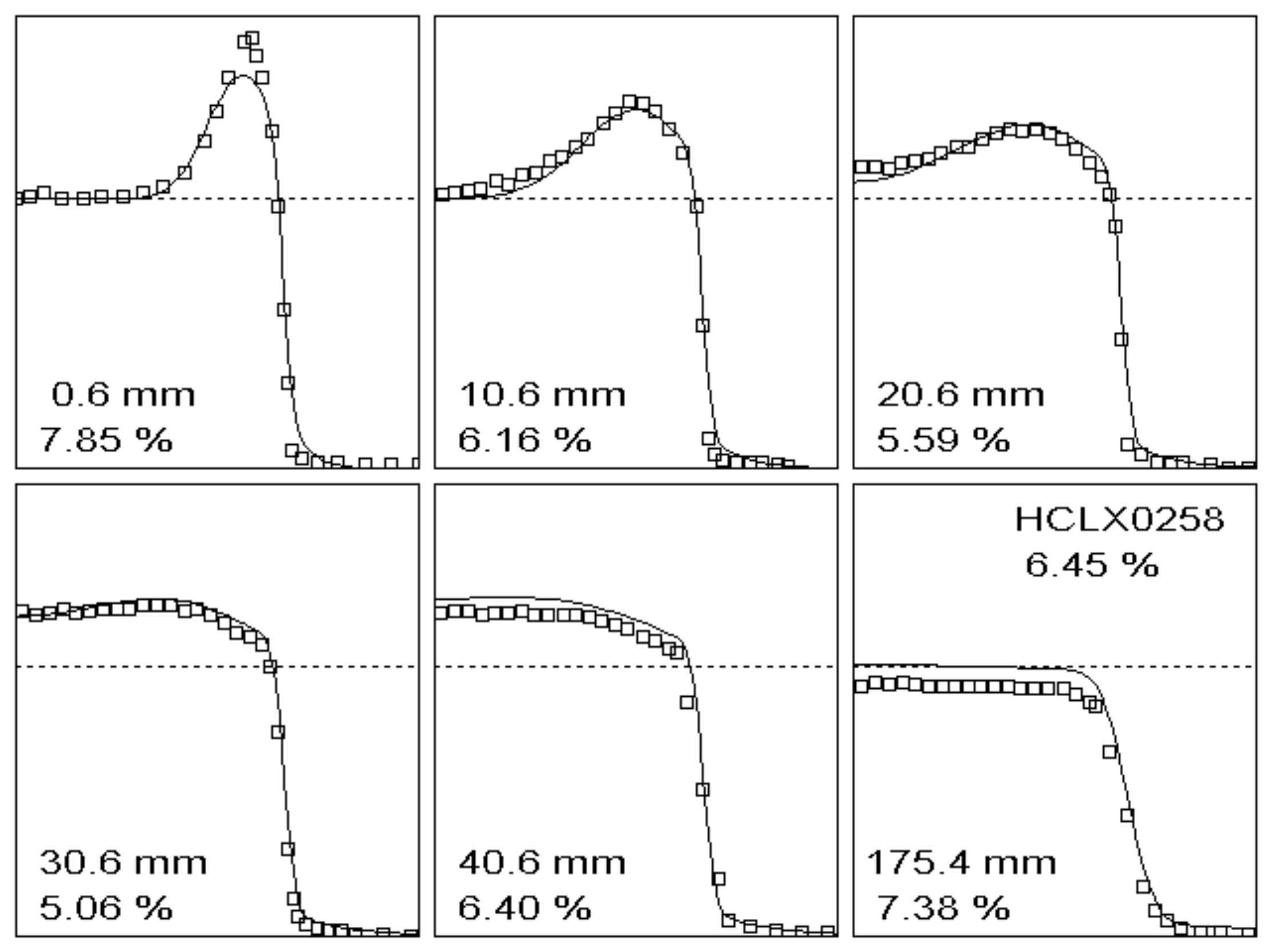}
  \caption{Transverse dose distributions in air at six distances (top number) from the downstream face of a brass collimator with a 9.88\,mm radius hole. Percentages give the rms deviation of the PBA calculation (line) from experiment (squares). The PBA calculation is normalized to the first experimental point in MP1 (0.6\,mm). The calculation took 1.4\,min. 1665\,K PBs were generated of which 531\,K traversed MP1-6.}
  \label{fig:HCLX0258}
\end{figure}

\begin{figure}[p] 
  \centering
  \includegraphics[bb=0 0 640 480,width=4.67in,height=3.5in,keepaspectratio]{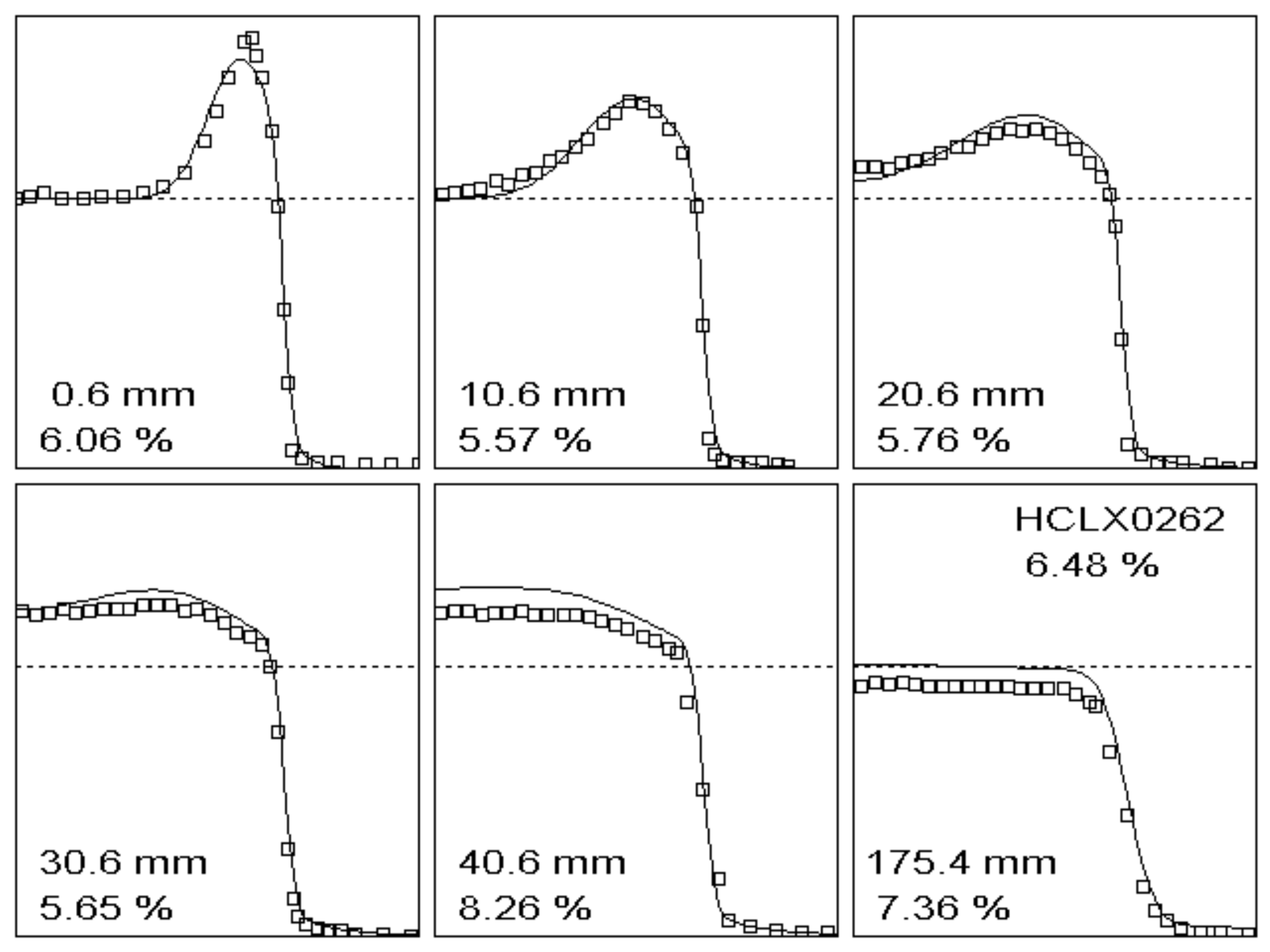}
  \caption{The same as Fig.\,\ref{fig:HCLX0258} except $p_3$ (Eq.\,\ref{eqn:splitTrigger}) was increased from 10 to 18. The calculation took 93\,min. 71\,M PBs were generated of which 40\,M traversed MP1-6. Agreement at MP1-2 is significantly better than Fig.\,\ref{fig:HCLX0258} but other MPs are the same or worse.}
  \label{fig:HCLX0262}
\end{figure}

\clearpage

\begin{figure}[p] 
  \centering
  \includegraphics[bb=0 0 640 480,width=4.67in,height=3.5in,keepaspectratio]{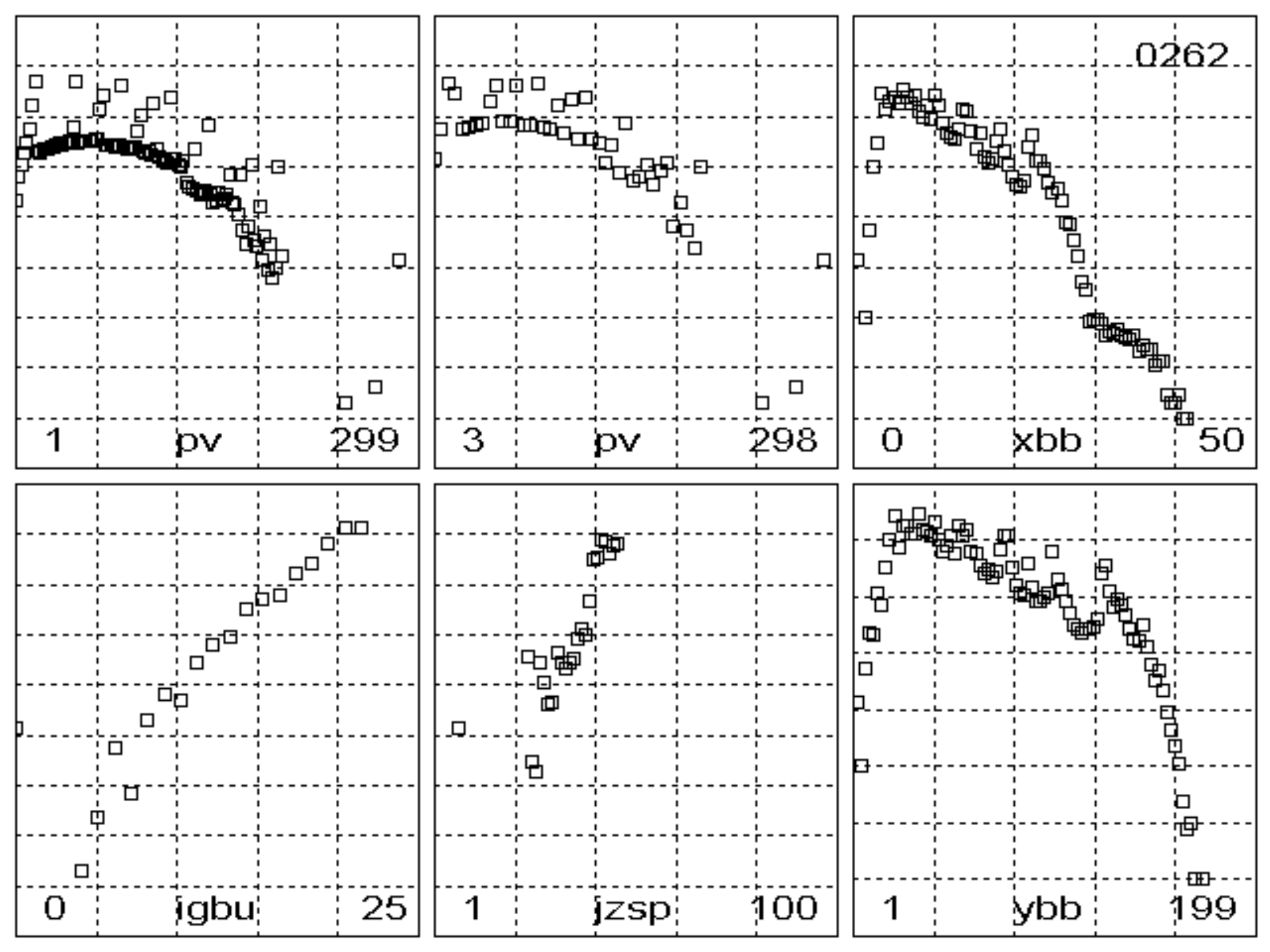}
  \caption{Frequency distributions for several quantities at MP6 in the PBA run corresponding to Fig.\,\ref{fig:HCLX0262}. The vertical scale is logarithmic by decades; the lowest horizontal line is at $10^0=1$. From top left: $pv$ (MeV), $pv$ (coarser bins), $\sigma_x$ (mm), generation index $i_g$, index $j_z$ of plane where split took place, $\sigma_\theta$ (mrad). The full range of each variable is indicated. Note non-physical spikes, particularly obvious in the $pv$ plots.}
  \label{fig:Logh0262}
\end{figure}

\begin{figure}[p] 
  \centering
  \includegraphics[bb=20 40 576 469,width=4.54in,height=3.50in,keepaspectratio]{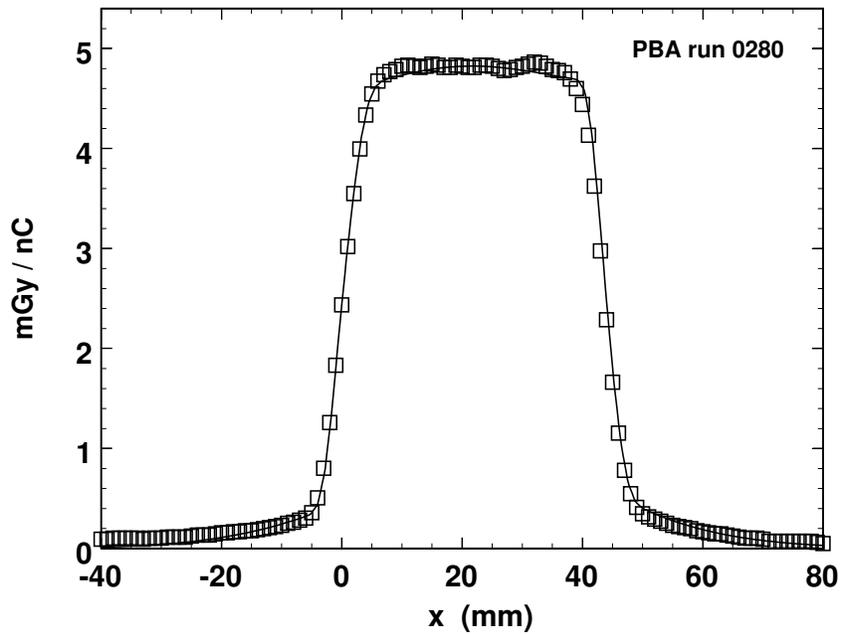}
  \caption{Dose distribution 200\,mm from downstream face of $40\times200$\,mm half-beam block in a 218\,MeV double-scattered beam. The line is the PBA calculation. The measurement (squares) is normalized and shifted but the $x$ scale, 1\,mm/point, is absolute. The L and R $80/20$ penumbras are 5.11, 4.65\,mm (PBA) and 5.26, 4.86\,mm (measured).}
  \label{fig:HBB0280}
\end{figure}

\clearpage

\begin{figure}[t]
\begin{verbatim}
HBBtest.INP :

run parameters:
MIXED  .01   1   0         range-energy; ztol, dpCprint, iReseed
218  42  218               incident MeV; MeV range for Overas fit
DOSE       0   999  .8     FLUENCE/DOSE; dose pv range; BP toe level
AIR        0   1.1  .5     S_EM dosetomatl, %/(g/cm2), %strag/R, %span
1000  3   0  1.15          S_EM nz,ns,za,zb/range
N   5   1   1              S_EM show, R/sig, zstep (mm), cut/R (%)
2.0  1.0   0    2   2      split if d<p1*sx & xbb>p2DBp3 & igbu<iglim(1,2)
VIR  .55  1.15             split source VIR,ESP,EFF; split sigx; spread
N  .5   1.15  0.00   3     redef1 show, sigx,spc,margin (mm); ruse/sigma
N  .5   1.15  2.00   3     redef2 ditto NB .5 <= spc <= 1.15mm
1    15  -9999             urbeam nRows, scanHW (mm), zSource (mm)
0   0    0    0            urbeam x0,y0,xp0,yp0 (mm,mrad)
0   0    0                 urbeam sigT,thtC,sigX (mr,mr,signed mm)
1  -40  80  101            dose #MPs, scan range, #pts
0  0  0  20  0  90         dose (x0,y0,phi)_1, (ditto)_2  (mm,mm,degr)
50  .05  .95               contours #, low,high level (relative)
0   0                      z1,z2 (mm) for terrain picture
terrain: mtl or file, #slabs, thickness (mm) - - - - - - - - - - - - - - -
LEAD         1       .875  (.9931/11.35 = .875)
AIR          1     18.747  (19.52 - .773)
CARBLD       1       .773  RM7 step1
LEAD         1      4.254  ditto
AIR          3    871.582  (917 - 4.254)
REDEF1
SS2.CON     48     56.
AIR          5   1412.164
REDEF2
HBB40.REC    3     60.
AIR          1    200.     downstr coll to isocenter
HISTOGRAMS
END         999    999     ================================================
- - - - - - - - - - - - - -  histogram requests  - - - - - - - - - - - - -
10  5  50000  50  3/  ndown, nacr, maxbuf, max #requests, max #cuts
'xco'   2   1 -100 200 1 1 1/
'igbu'  7   1 -.5  20  1/
'yco'   3   2 -200 200 1 1 3/
'igbu'  7   1 -.5  20  1/
'pv'    6   2  0  200  1  1 5  340  200/
'pv'    6   8  0   50  1  2 5  340  200/
'xbb'  10   1  0  100  1  3 5/
'igbu'  7   1 -.5  20  1  4 5/
'jzsp'  8   1  50  50  1  5 5/
'ybb'  11   2  0  100  1  6 5/ 
'end'/

HBB40.REC :

BRASS               universe
AIR                 inner
0 -100  40  100     LL,UR (mm)
\end{verbatim}
\caption{Input files for the half-beam block example.\label{fig:HBBfiles}}
\end{figure}

\clearpage

\begin{figure}[p] 
  \centering
  \includegraphics[bb=0 0 552 503,width=4.94in,height=4.5in,keepaspectratio]{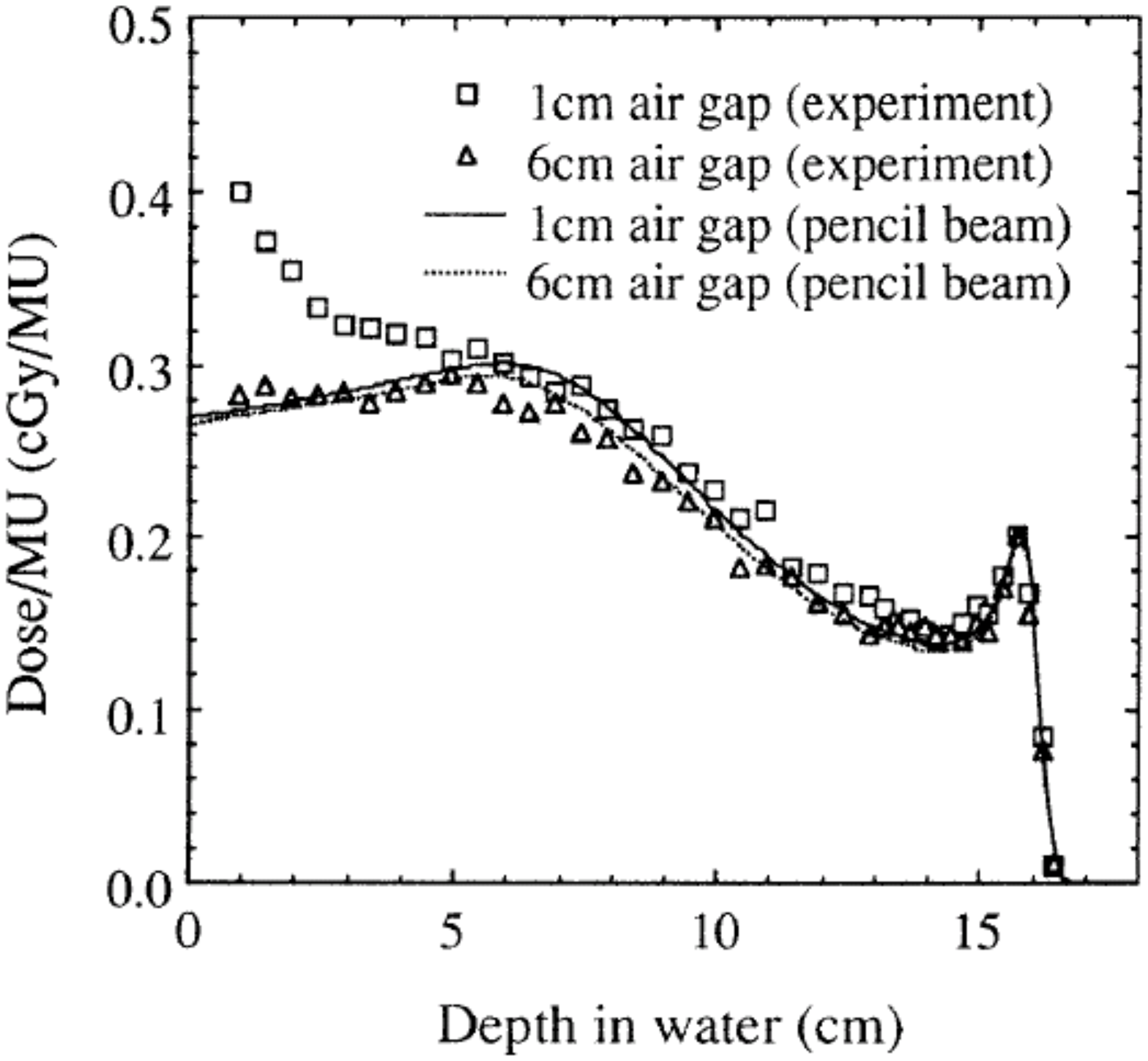}
  \caption{Original Fig.\,11 of Hong et al. \cite{hong}.}
  \label{fig:urHong11}
\end{figure}

\begin{figure}[p] 
  \centering
  \includegraphics[bb=0 0 640 480,width=4.67in,height=3.5in,keepaspectratio]{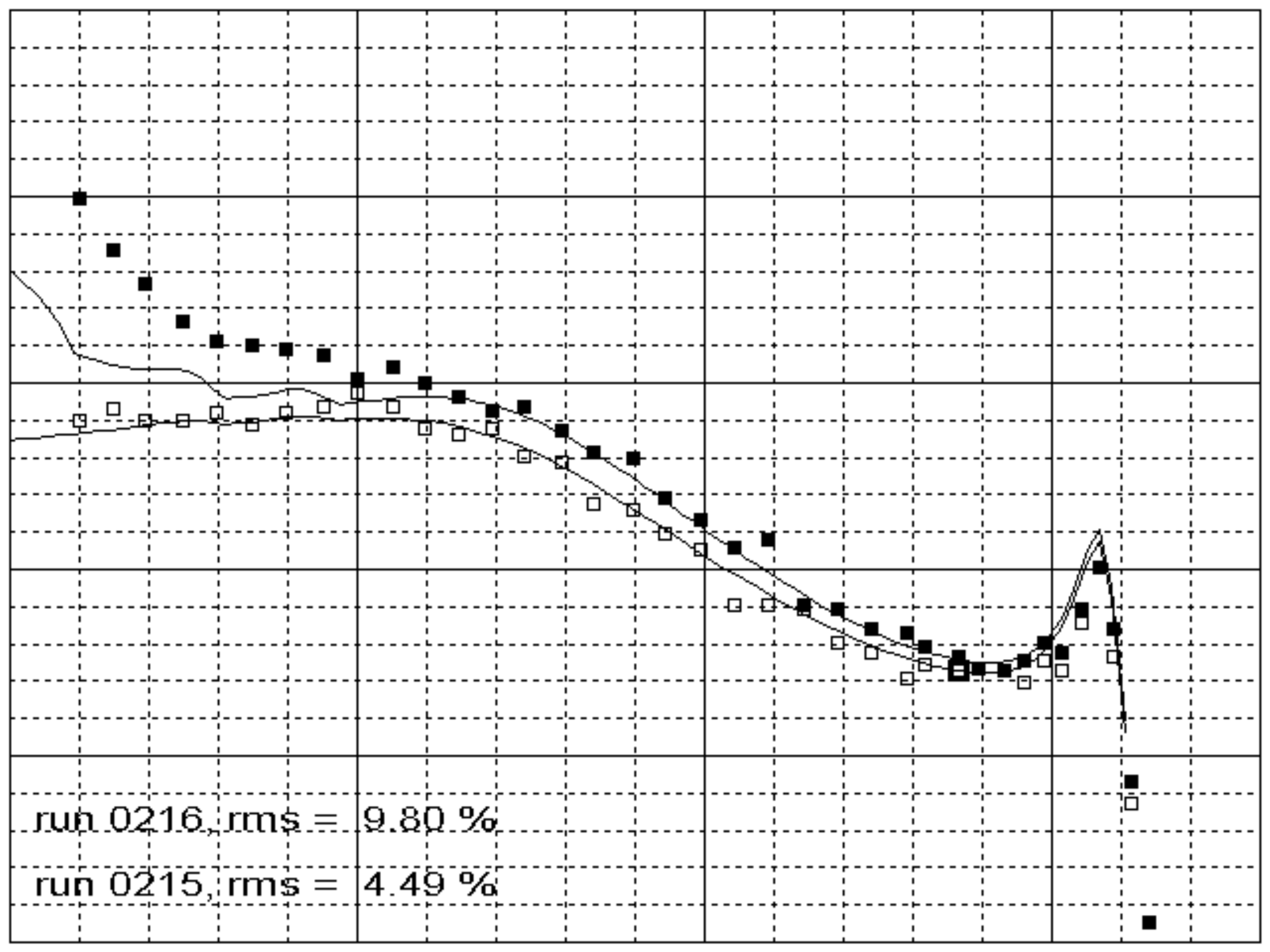}
  \caption{PBA computation corresponding to Hong et al. \cite{hong} Fig.\,11 (see Fig\,\ref{fig:urHong11} for key).}
  \label{fig:Hong11}
\end{figure}

\begin{figure}[p] 
  \centering
  \includegraphics[bb=0 0 791 713,width=4.99in,height=4.5in,keepaspectratio]{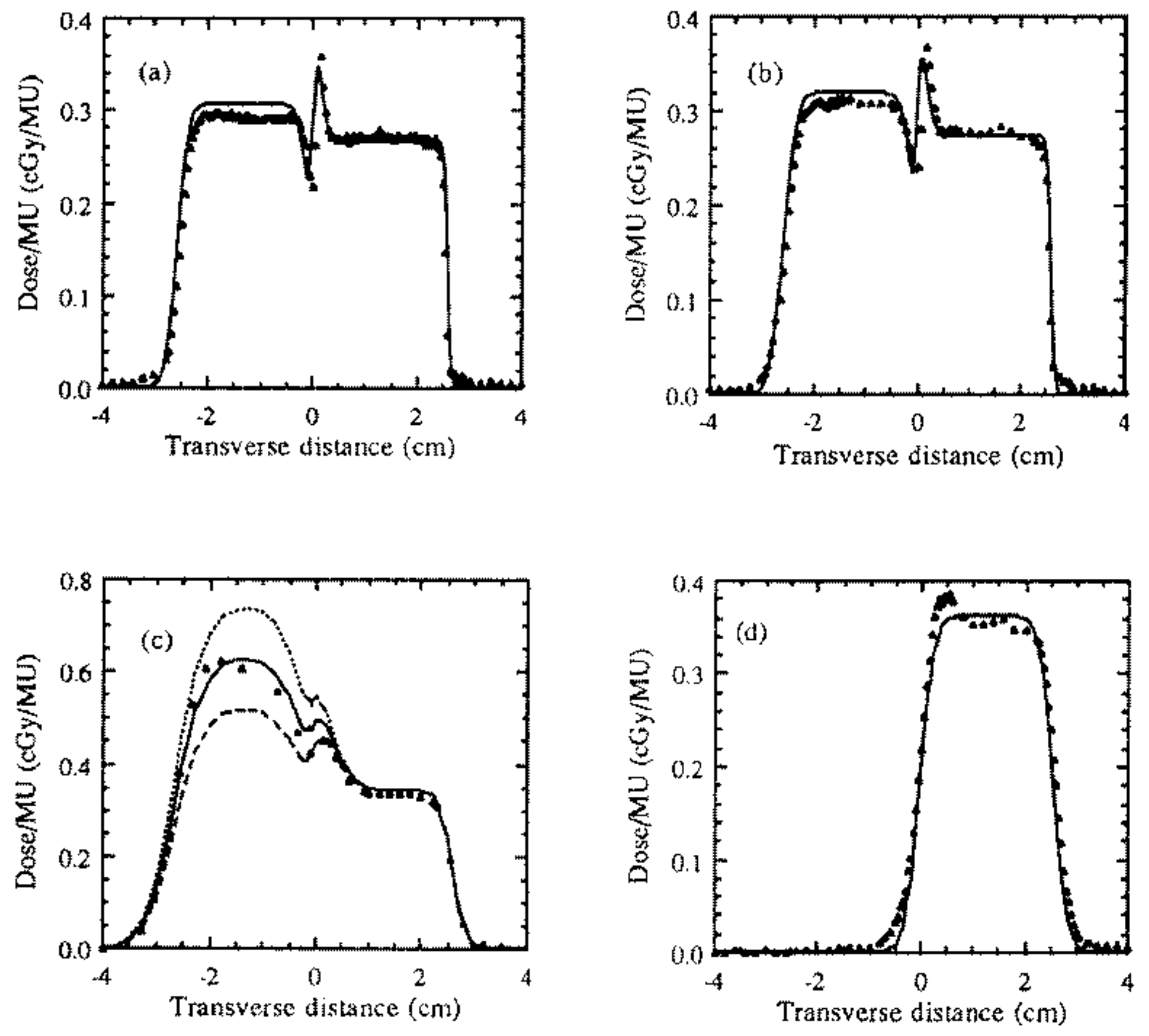}
  \caption{Original Fig.\,14 of Hong et al. \cite{hong}.}
  \label{fig:urHong14}
\end{figure}

\begin{figure}[p] 
  \centering
  \includegraphics[bb=0 0 640 480,width=4.67in,height=3.5in,keepaspectratio]{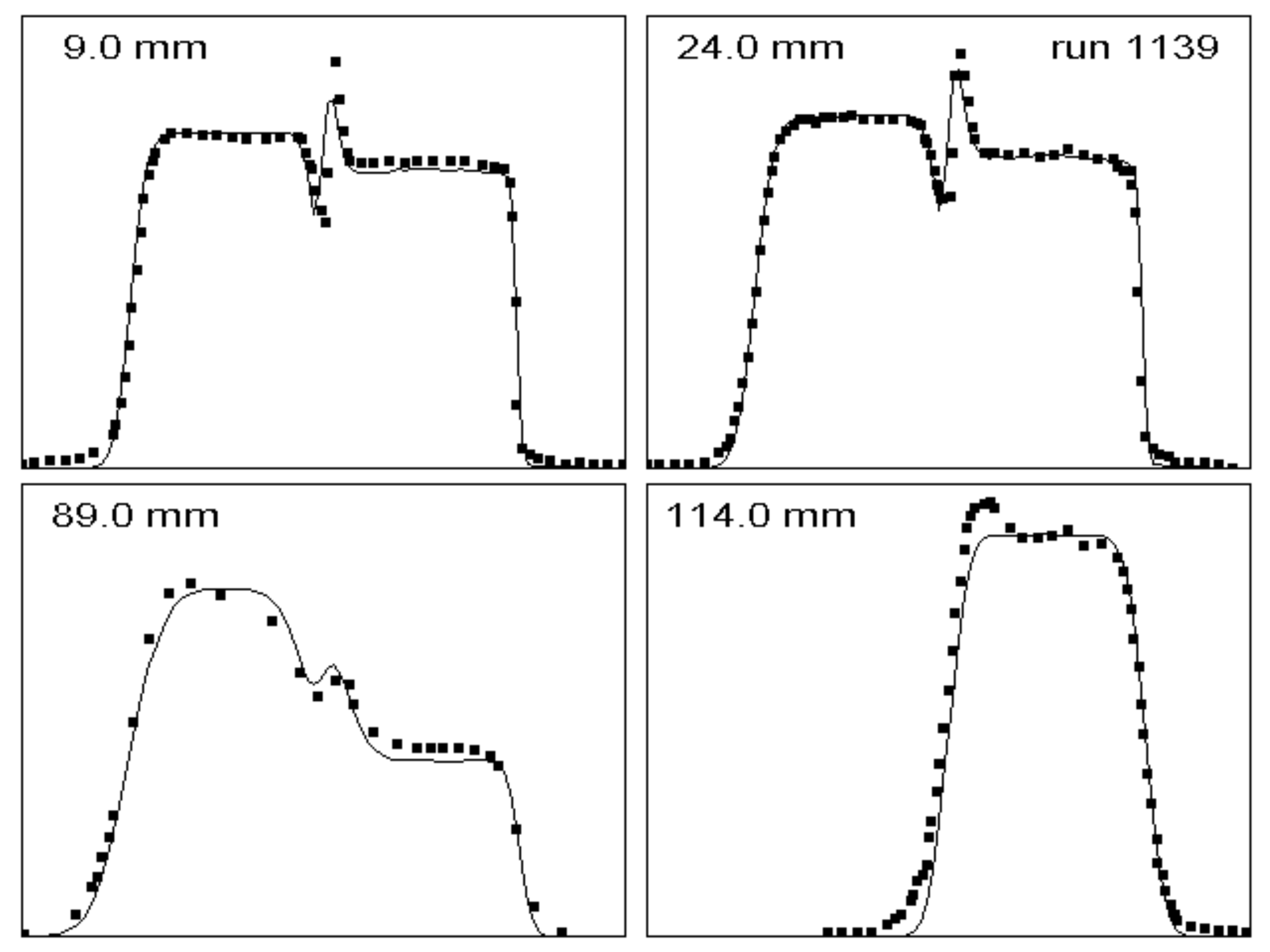}
  \caption{PBA computation corresponding to Fig.\,14 of Hong et al. \cite{hong}.}
  \label{fig:Hong14}
\end{figure}

\clearpage

\begin{figure}[p] 
  \centering
  \includegraphics[bb=0 0 640 480,width=4.54in,height=3.5in,keepaspectratio]{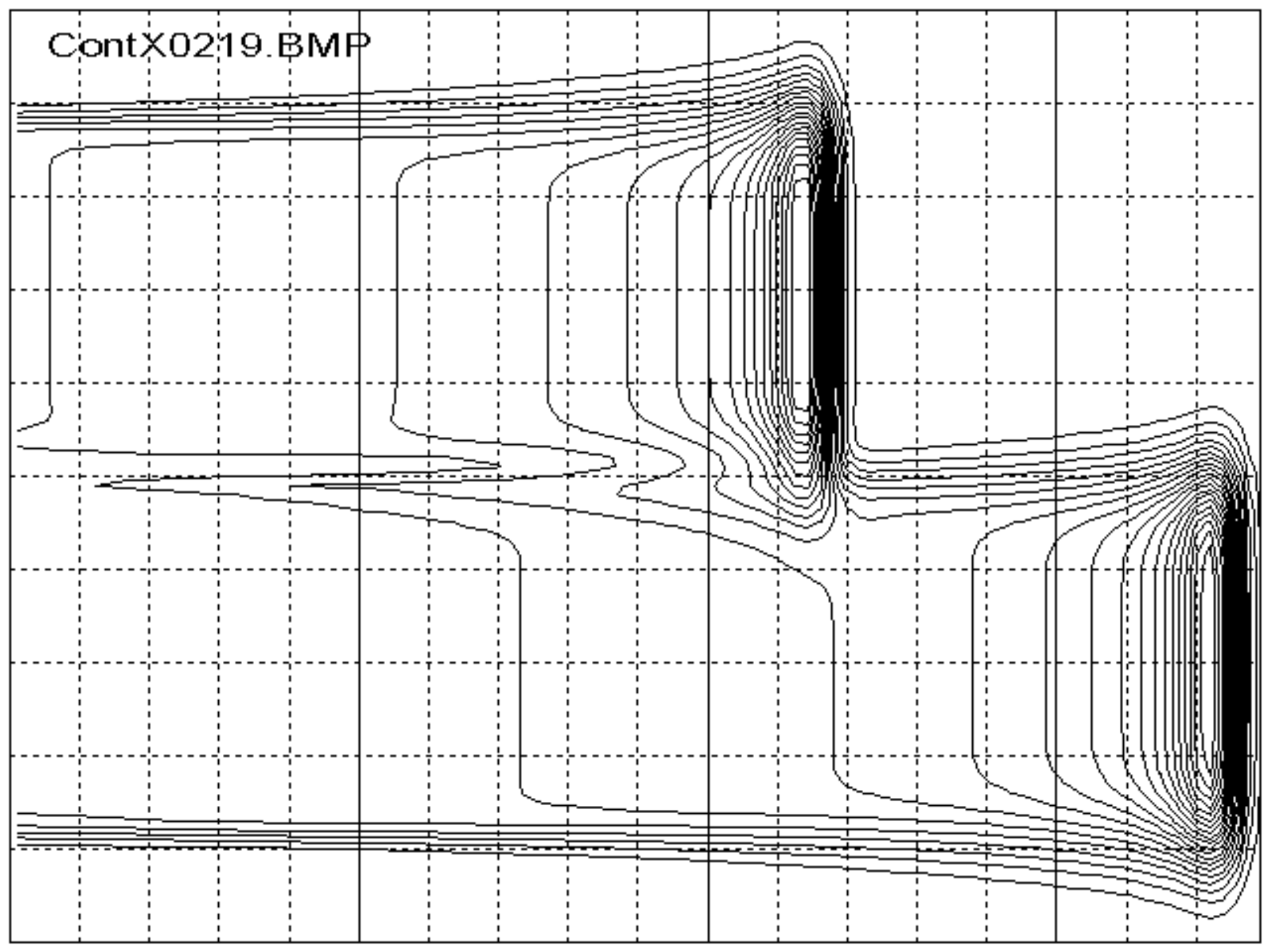}
  \caption{Contour plot for 5\,cm Lucite in front of water tank.}
  \label{fig:ContLucite}
\end{figure}

\end{document}